\documentclass[twocolumn,pre,amsmath,amssymb,superscriptaddress,notitlepage,footinbib]{revtex4-1}

\usepackage{braket}
\usepackage{amsmath}
\usepackage{amsfonts}
\usepackage{xcolor}
\usepackage{amssymb}
\usepackage{graphicx}
\usepackage[caption=false]{subfig}
\usepackage{natbib}

\begin{document}
\title{Strong coupling in thermoelectric nanojunctions: a reaction coordinate framework}
\author{Conor McConnell}
\email{conor.mcconnell@postgrad.manchester.ac.uk}
\affiliation{Department of Physics and Astronomy, University of Manchester, Oxford Road, Manchester M13 9PL, United Kingdom}
\author{Ahsan Nazir}\email{ahsan.nazir@manchester.ac.uk}
\affiliation{Department of Physics and Astronomy, University of Manchester, Oxford Road, Manchester M13 9PL, United Kingdom}
\date{\today}

\begin{abstract}
We study %consider 
a model of a thermoelectric 
nanojunction driven by %exhibiting %strong 
vibrationally-assisted %phonon-assisted 
tunneling. %electron-vibrational coupling 
We apply the reaction coordinate formalism to derive %various
a master equation governing its thermoelectric performance %behaviour 
beyond the weak electron-vibrational %vibrational 
coupling limit. %We focus on thermoelectric regimes of performance, 
Employing %both particle and energy 
full counting statistics we %to 
calculate the current flow, thermopower, associated noise, and efficiency %of our model thermoelectric 
without resorting to the weak vibrational coupling approximation. We demonstrate intricacies of the power-efficiency-precision trade-off at strong coupling, showing %definitively 
that the three cannot be maximised simultaneously in our model. 
Finally, we emphasise the importance of capturing non-additivity %treatment 
when considering %various
strong coupling and multiple environments, demonstrating that an additive treatment of the environments can violate the upper bound on thermoelectric efficiency imposed by Carnot.
\end{abstract}

\maketitle
\section{Introduction}
The understanding of quantum transport and quantum thermodynamics is at the forefront of designing practical quantum devices. The study of %thermoelectric 
molecular and solid-state nanojunctions combines these two areas, 
with the aim of %realisation of 
realising functional quantum circuitry \cite{Aviram74,Reed252,Dekker97,Park99,Nicewarner01,Cui01,Nitzan03,Kubatkin03,Tao06, Chen07, Aradhya13, Sun14, Nichols15,Nichols15}. %Molecular 
Even minimal nanojunction %can be difficult to model, however, due to 
models often require the inclusion of several different environments, which also presents an interesting theoretical challenge. Typically, these environments represent two key features of a quantum circuit: the electronic leads, which are modelled as fermionic reservoirs; and the vibrational effects of the molecule or solid-state lattice, modelled by %considered as 
a reservoir of bosonic modes. %(potentially strongly-coupled) bosonic bath. %\par

For both molecular and solid-state nanojunctions understanding the influence of electron-vibrational coupling on transport properties is of crucial importance, and has been studied widely (see for example~\cite{Kushmerick04,Koch05,Koch205,Galperin06,PhysRevB.74.205438,Brandes07,Brandes09,Thoss09,PhysRevB.82.115314,PhysRevLett.107.046802,Thoss11,PhysRevB.85.075412,C2CP40851A,Santamore13,PhysRevB.87.085422,Schaller13,Schaller14,PhysRevB.92.245418,Strasberg16,Pigeon17,Stones17,Sowa17,Sowa217,McConnell19,PhysRevB.101.075422}). In particular, strong vibrational coupling can result in striking qualitatively changes to transport behaviour, such as the suppression of low-bias current known as Franck-Condon blockade~\cite{Koch05,PhysRevB.74.205438}.  

In this work, rather than focussing on electron transport %down %along %driven by 
%a chemical potential gradient 
driven by a voltage bias, we study the influence of strong vibrational coupling on the thermoelectric performance of a model nanojunction, where current is driven against a bias by a temperature gradient. %As such 
The system thus generates a finite %work or 
power output with an associated efficiency, determined also by the heat flux. We are particularly interested in understanding the interplay between the power, efficiency, %as well as other metrics such as 
and noise, which is of key importance in assessing the performance of %quantum transport literature at present 
thermodynamic devices~\cite{PhysRevE.93.052145,PhysRevLett.116.120601,Seifert18,Liu19,Saryal19,Goold19,PhysRevLett.125.260604,PhysRevLett.126.210603}.

We consider a two-site nanojunction model where inter-site transport is mediated via phonon-assisted tunneling (PAT). This is a natural setting to study the impact of vibrational coupling in a thermoelectric context as phonons are then directly responsible for promoting electron transport. To model the system at strong electron-phonon coupling we %are going to make use of a 
employ the reaction coordinate master equation (RCME) approach~\cite{Strasberg16,IlesSmith14,IlesSmith16,Maguire18,RCReview,McConnell19}. %To do so we 
This involves applying a unitary Hamiltonian-level mapping to incorporate a single collective mode from the phonon environment into our electronic system, generating what we shall term an augmented system. The remaining %residual 
phonon modes are then weakly coupled to %this single collective mode 
the augmented system and traced out within the standard Born-Markov (BM) approximations, though retaining %In this way we can capture 
the dominant strong-coupling and non-Markovian effects on the original electronic system. Several other theoretical techniques have been applied to related models of vibrationally coupled charge transport, including investigations of thermoelectric performance. Examples include Green's function approaches~\cite{PhysRevB.82.115314,PhysRevLett.107.046802,PhysRevB.87.085422,PhysRevB.92.245418}, linear response methods~\cite{PhysRevB.85.075412}, and full counting statistics with the hierarchical equations of motion~\cite{PhysRevB.101.075422} and quantum master equations~\cite{C2CP40851A,PhysRevB.92.245418}, where strong lead couplings have also been considered. The impact of strong light-matter (rather than vibrational) coupling in thermoelectric devices has also been explored~\cite{PhysRevB.99.035129}.

One significant advantage of the RCME approach is that it allows for a straightforward incorporation of non-additive effects between the strongly-coupled phonons and the fermionic leads, by deriving the lead influences within the full eigenbasis of the augmented system. This is crucial for studying thermoelectric systems since they operate at finite bias, and an additive treatment in which the leads are insensitive to the strong system-phonon coupling 
can only be justified in the infinite bias limit~\cite{McConnell19}. In fact, our non-additive and additive comparisons %important as it shows
show directly that the additive approximation can lead to unphysical artefacts in thermoelectric performance no matter how sophisticated the treatment of phonons, such as thermoelectric efficiencies beyond the Carnot bound. 

We shall also compare the RCME treatment to a weak coupling master equation (WCME) that treats all environments, including the phonons, within the standard BM approximations. This confirms consistency between all methods at weak enough phonon coupling strength, and highlights the limitations of the WCME as the phonon coupling is increased. For all methods we employ counting statistics techniques \cite{Koch05, Schaller13, Taniguchi20, Schaller14, Flindt05, Flindt10} to calculate the transport cumulants necessary to evaluate the thermoelectric power output, noise, and efficiency. 

The paper is organised as follows. In Sec.~\ref{model} we outline the theoretical model and techniques used, 
including the WCME, the reaction-coordinate (RC) mapping, and 
the implementation of counting statistics. In Sec.~\ref{leftresource} we %shall set up a 
consider a thermoelectric regime in which a temperature gradient is applied across the leads, with the left lead acting as a resource such that the phonons not only mediate transport but also act as a heat loss mechanism. We contrast this with the case of a phonon resource 
in Sec.~\ref{phononresource}. Finally, we summarise our results and discuss future directions in Sec.~\ref{conc}.

\begin{figure}
\center
\includegraphics[width=0.75\linewidth]{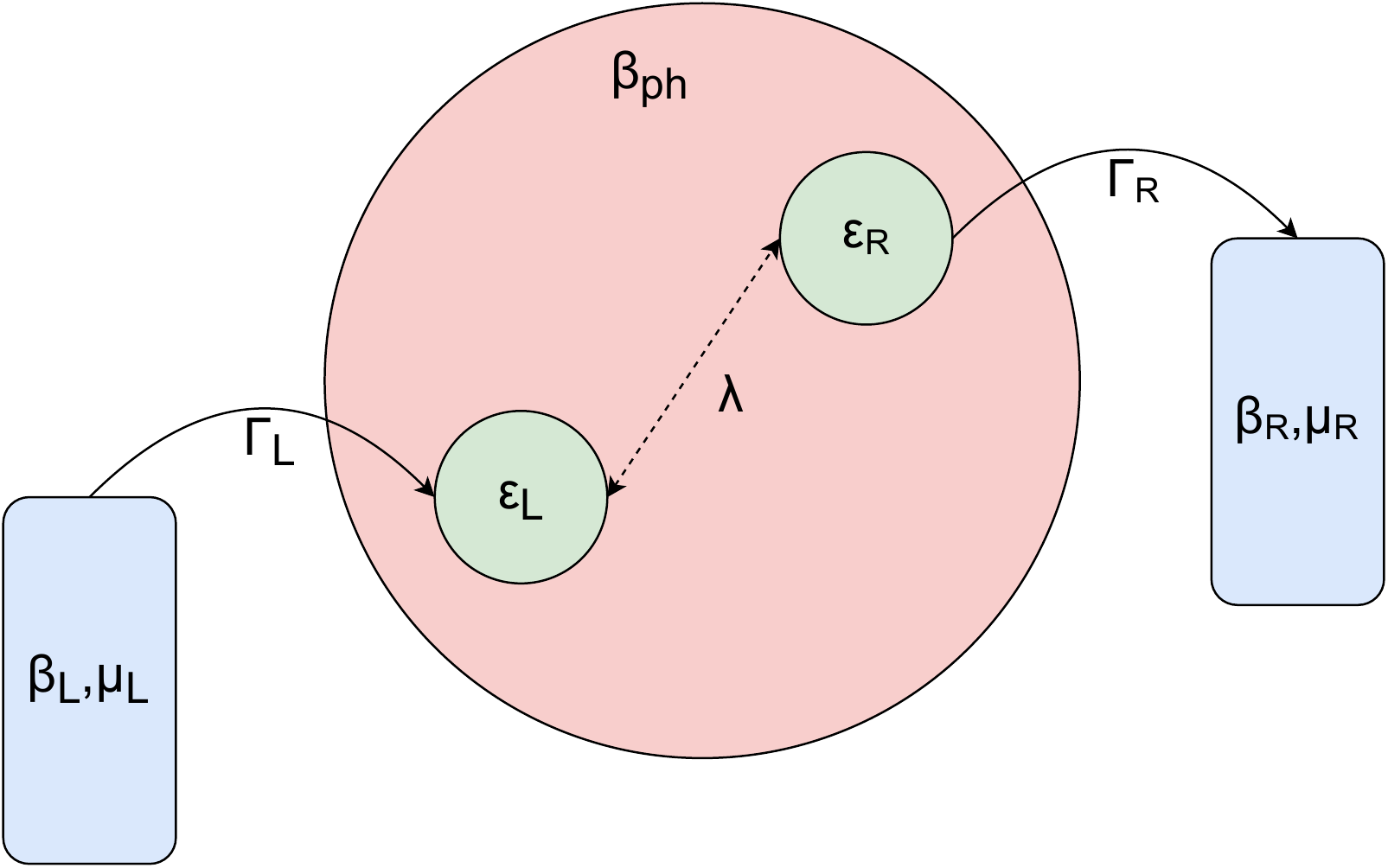}
\caption{Schematic diagram of the two-site thermoelectric model. %with phonon-assisted tunnelling, acting in a thermoelectric regime. 
Each site is connected to an individual lead and both sites are coupled through %$\lambda$ to 
a common phonon bath at inverse temperature $\beta_{\rm ph}$, with reorganisation energy $\lambda$. The site energies are given by $\epsilon_{\rm i}$, %$\epsilon_{L/R}$ gives the energy of each dot, while $\lambda$ and 
couplings to the leads are denoted $\Gamma_{\rm i}$, %and $\Gamma_{\rm R}$, respectively,
and each lead is held at a chemical potential $\mu_{\rm i}$ and inverse temperature $\beta_{\rm i}$, where ${\rm i}={\rm L},{\rm R}$.}
\label{fig:PATDiagram}
\end{figure}

\section{Theoretical Model}\label{model}

Our two-site model is shown schematically in Fig.~\ref{fig:PATDiagram}. 
Each site is coupled individually to a fermionic lead and both are coupled to a common phonon bath that mediates electron transport. 
This could describe, for example, a molecule or double quantum dot with two regions of high electron density (the dots), coupled via 
molecular vibrations or the solid-state lattice. %are modelled as a phonon bath. 
Direct inter-site tunneling can also be included straightforwardly, but to isolate the impact of vibrations we shall not do so here.

We consider the left lead to be held at a lower chemical potential ($\mu_{\rm L}$) than the right ($\mu_{\rm R}$). In order for electrons to overcome the chemical potential gradient, %and a current to
leading to a useful current flow, we require a temperature gradient. In the first instance (Sec.~\ref{leftresource}) we set the left lead temperature to be greater than that of the right lead or phonons, %$T_{\rm L}>T_{\rm R},T_{\rm ph}$, 
implying $\beta_{\rm L}<\beta_{\rm R}=\beta_{\rm ph}$ for the inverse temperatures. Subsequently, in Sec.~\ref{phononresource}, we shall set the phonon temperature to be greater than that of both leads, which we fix equal ($\beta_{\rm L}=\beta_{\rm R}=\beta_{\rm el}$), so that $\beta_{\rm ph}<\beta_{\rm el}$ and the phonons become the resource. 
The right dot is also held at a higher energy than the right lead to promote electron flow from left to right and suppress the back-flow of electrons. As a consequence some of the work done to push the electrons onto the right dot is lost as heat when they are transferred to the right lead. However, as $\mu_{\rm R}>\mu_{\rm L}$ we still have a positive work output overall.

The Hamiltonian for our set-up is 
\begin{equation}
H=H_S+H_{E_{\rm L}}+H_{E_{\rm R}}+H_{E_{\rm ph}}+H_{I_{\rm L}}+H_{I_{\rm R}}+H_{I_ {\rm ph}},
\end{equation} 
with system Hamiltonian 
\begin{equation}\label{Hsys}
H_S=\epsilon_{\rm L} d_{\rm L}d_{\rm L}^{\dagger} + \epsilon_{\rm R} d_{\rm R}d_{\rm R}^{\dagger} +Ud_{\rm L}d_{\rm L}^{\dagger}d_{\rm R}d_{\rm R}^{\dagger},
\end{equation}
where $ d_{\rm L}$ and $ d_{\rm R}$ are fermionic annihilation operators for the left and right dot, respectively, $\epsilon_{\rm L}$ ($\epsilon_{\rm R}$) is the energy of the left (right) dot, and $U$ is the Coulomb energy. We assume that on-site Coulomb repulsion is large such that each dot is never occupied by more than a single excess electron. 
The terms corresponding to the fermionic leads are
\begin{align}
H_{E_{\rm L}}&= \sum_k \epsilon_{k_{\rm L}}c^\dagger_{k_{\rm L}}c_{k_{\rm L}},\\
H_{E_{\rm R}}&= \sum_k \epsilon_{k_{\rm R}}c^\dagger_{k_{\rm R}}c_{k_{\rm R}},
\end{align}
and their interactions with the system are given by
\begin{align}
H_{I_{\rm L}}&= \sum_k (t_{k_{\rm L}}d_{\rm L}c^\dagger_{k_{\rm L}}+t_{k_{\rm L}}^*c_{k_{\rm L}}d_{\rm L}^{\dagger}),\\
H_{I_{\rm R}}&=\sum_k (t_{k_{\rm R}}d_{\rm R}c_{k_{\rm R}}^\dagger+t_{k_{\rm R}}^*c_{k_{\rm R}}d_{\rm R}^{\dagger}).
\end{align}
Here $\epsilon_{k_{\rm i}}$ is the energy of the $k^{th}$ fermionic mode in lead ${\rm i}$, which is coupled to the system via $t_{k_{\rm i}}$, and $c_{k_{\rm i}}$ are fermionic annihilation operators for electrons in the leads. 

As the system and lead fermionic operators anti-commute, the interaction Hamiltonians $H_{I_{\rm L}}$ and $H_{I_{\rm R}}$ do not have the tensor product structure required for our subsequent master equation derivations. However, we can use a Jordan-Wigner transformation to rectify this~\cite{Schaller14}, after which the state space for the central system can be written \{$\ket{G},\ket{L},\ket{R},\ket{D}$\}. Here $\ket{G}$ is the ground state, with neither dot occupied by an excess electron, $\ket{L}$ ($\ket{R}$) denotes an electron present the on the left (right) dot, and $\ket{D}$ is the state in which both dots are occupied. 
%We assume that on-site Coulomb repulsion is large such that each dot is never occupied by more than a single excess electron.
%The Hamiltonian for our set-up is 
%\begin{equation}
%H=H_S+H_{E_{\rm L}}+H_{E_{\rm R}}+H_{E_{\rm ph}}+H_{I_{\rm L}}+H_{I_{\rm R}}+H_{I_ {\rm ph}},
%\end{equation} 
%with %system %Hamiltonian 
Then, the system and interaction Hamiltonians can be written as
\begin{equation}\label{Hsys}
H_S=\epsilon_{\rm L} \ket{L}\bra{L} + \epsilon_{\rm R} \ket{R}\bra{R} +(\epsilon_{\rm L} + \epsilon_{\rm R}+ U)\ket{D}\bra{D},
\end{equation}
%where $\epsilon_{\rm L}$ ($\epsilon_{\rm R}$) is the energy of the left (right) dot and $U$ is the Coulomb energy. 
%The terms corresponding to the fermionic leads are
%\begin{align}
%H_{E_{\rm L}}&= \sum_k \epsilon_{k_{\rm L}}c^\dagger_{k_{\rm L}}c_{k_{\rm L}},\\
%H_{E_{\rm R}}&= \sum_k \epsilon_{k_{\rm R}}c^\dagger_{k_{\rm R}}c_{k_{\rm R}},
%\end{align}
%and their interactions with the system are given by
and
\begin{align}
H_{I_{\rm L}}&= (\ket{R}\bra{D}-\ket{G}\bra{L})\sum_k t_{k_{\rm L}}c^\dagger_{k_{\rm L}}
\nonumber\\&+(\ket{D}\bra{R}-\ket{L}\bra{G})\sum_k t_{k_{\rm L}}^*c_{k_{\rm L}},\\
H_{I_{\rm R}}&=(\ket{L}\bra{D}+\ket{G}\bra{R})\sum_k t_{k_{\rm R}}c_{k_{\rm R}}^\dagger
\nonumber\\&+(\ket{D}\bra{L}+\ket{R}\bra{G})\sum_k t_{k_{\rm R}}^*c_{k_{\rm R}},
\end{align}
respectively, 
%Here $\epsilon_{k_{\rm i}}$ is the energy of the $k^{th}$ fermionic mode in lead ${\rm i}$, which is coupled to the system via $t_{k_{\rm i}}$, and $c_{k_{\rm i}} (c^{\dagger}_{k_{\rm i}})$ are standard fermionic annihilation (creation) operators. 
where the negative terms in the left lead couplings %are negative to 
preserve the system fermionic anti-commutation relations after the Jordan-Wigner transformation. %of fermions.\par

We consider a phonon environment coupled to the system in Brownian oscillator form~\cite{Strasberg16, PhysRevA.37.4419, Weiss12}
\begin{align}
H_{E_{\rm ph}}+H_{I_{\rm ph}}&=\frac{1}{2}\sum_{q} \Bigg[p_q^2+\omega_q^2\left(x_q-\frac{c_q}{\omega_q^2}s\right)^2\Bigg], 
\end{align} 
with system coupling operator $s=d_{\rm L}d_{\rm R}^{\dagger}+d_{\rm R}d_{\rm L}^{\dagger}$. Here, $x_q$ and $p_q$ are phonon position and momentum operators satisfying $[x_q,p_{q'}]=i\delta_{qq'}$ ($\hbar$ is set to $1$ throughout), $\omega_q$ is the frequency of the $q^{th}$ phonon mode, which couples to the system with strength $c_q$. Expressing the phonon position and momentum operators in terms of standard bosonic creation and annihilation operators $a_q^{\dagger}$ and $a_q$, the phonon environment Hamiltonian becomes (ignoring a constant term) 
\begin{align}
H_{E_{\rm ph}}&=\sum_{q} \omega_q a_q^\dagger a_q.
\end{align} 
Likewise, after applying the Jordan-Wigner transformation the interaction Hamiltonian is given by
\begin{align}\label{HIph}
H_{I_{\rm ph}}&=(\ket{L}\bra{R}+\ket{R}\bra{L})\sum_q h_q(a_q^\dagger + a_q) \nonumber\\&+ \sum_q\frac{h_q^2}{\omega_q}(\ket{L}\bra{R}+\ket{R}\bra{L})^2,
\end{align}
with $h_q=-c_q/\sqrt{2\omega_q}$.
%Here $\omega_q$ is the frequency of the $q^{th}$ phonon mode, with coupling $h_q$, and $a_q (a_q^{\dagger})$ are standard bosonic annihilation (creation) operators. 
The final term in Eq.~(\ref{HIph}) 
compensates for energetic renormalisation of the system caused by the phonon coupling~\cite{Weiss12} and ensures that the sites remain within the lead bias window even for strong phonon interactions. 

\subsection{Weak coupling master equation}

For sufficiently weak coupling between the system and %the various 
environments we expect to be able to apply the standard BM approximations to arrive at a Redfield master equation~\cite{Breuer:2002aa}. This treats the environments additively, meaning that each environment influences the system in the same manner as it would if treated in isolation.

We write the lead interaction Hamiltonians as %where we have
\begin{equation}
H_{I_{\rm L}}=A_1 B_1 +A_2  B_2,
\end{equation}
and
\begin{equation}
H_{I_{\rm R}}=A_3 B_3 +A_4  B_4,
\end{equation}
with $A_1=-\ket{G}\bra{L}+\ket{R}\bra{D}$, $A_2=-\ket{L}\bra{G} + \ket{D}\bra{R}$, $A_3=\ket{G}\bra{R}+\ket{L}\bra{D}$, 
$A_4=\ket{R}\bra{G} + \ket{D}\bra{L}$, $B_1= \sum_k t_{k_{\rm L}}c^\dagger_{k_{\rm L}}$, $B_2=\sum_k t_{k_{\rm L}}^* c_{k_{\rm L}}$, $B_3= \sum_k t_{k_{\rm R}}c^\dagger_{k_{\rm R}}$ and $B_4=\sum_k t_{k_{\rm R}}^* c_{k_{\rm R}}$. Considering $H_{I_{\rm L}}$ first, and following the textbook BM procedure, we arrive at the following Liouvillion for the left lead contribution to the dynamics of the reduced system density operator $\rho_S(t)$~\cite{Breuer:2002aa},
\begin{align}\label{LeftLeadMEPAT}
\mathcal{L}_{\mathrm{L}}{\rho}_S(t)=&-\int^\infty_0 d\tau [A_1,A_2(-\tau)\rho_S(t)]C_{12}(\tau)\nonumber\\&+[\rho_S(t)A_1(-\tau),A_2]C_{12}(-\tau)\nonumber\\&+[A_2,A_1(-\tau)\rho_S(t)]C_{21}(\tau)\nonumber\\&+[\rho_S(t)A_2(-\tau),A_1]C_{21}(-\tau).
\end{align}
Here, the time-dependent 
system operators and bath correlation functions are given by %defined as
\begin{align}
A_1(t)&=-\ket{G}\bra{L} e^{i \epsilon_{\rm L} t}+\ket{R}\bra{D} e^{i (\epsilon_{\rm L} +U) t},\\
A_2(t)&=-\ket{L}\bra{G} e^{-i \epsilon_{\rm L} t} + \ket{D}\bra{R}  e^{-i (\epsilon_{\rm L} +U) t},\\
C_{12}(t)&=\mathrm{tr}_{E}(e^{iH_{E_{\rm L}}t}B_{1}e^{-iH_{E_{\rm L}}t}B_2 \rho_{\rm L})\nonumber \\&=\sum_{k_{\rm L}}|t_{k_{\rm L}}|^2 f_{\rm L}(\epsilon_{k_{\rm L}}-\mu_{\rm L})e^{i \epsilon_{k_{\rm L}}t},\\
C_{21}(t)&=\mathrm{tr}_{E}(e^{iH_{E_{\rm L}}t}B_{2}e^{-iH_{E_{\rm L}}t}B_1 \rho_{\rm L})\nonumber \\&=\sum_{k_{\rm L}}|t_{k_{\rm L}}|^2 (1-f_{\rm L}(\epsilon_{k_{\rm L}}-\mu_{\rm L}))e^{-i \epsilon_{k_{\rm L}}t},
\end{align}
where %$B_{m}(t)=e^{iH_{E_{\rm L}}t}B_{m}e^{-iH_{E_{\rm L}}t}$ and 
$\rho_{\rm L}={e^{-\beta_{\rm L} (H_{E_{\rm L}}-\mu_{\rm L}N_{E_{\rm L}})}}/{\mathrm{tr}[e^{-\beta_{\rm L} (H_{E_{\rm L}}-\mu_{\rm L}N_{E_{\rm L}})}]}$ is the reduced density operator %describing the environmental degrees of freedom 
of the left lead, which we have taken to be in equilibrium. 
We also define $N_{E_{\rm i}}=\sum_k c^\dagger_{k_{\rm i}}c_{k_{\rm i}}$ and the  
Fermi-Dirac distribution $f_{\rm i}(\omega)=(e^{\beta_{\rm i} (\omega -\mu_{\rm i})}+1)^{-1}$, where $\beta_{\rm i}$ is the inverse temperature of fermionic environment ${\rm i}={\rm L},{\rm R}$ and $\mu_{\rm i}$ its chemical potential. 

The right lead contribution follows in complete analogy, 
\begin{align}\label{RightLeadMEPAT}
\mathcal{L}_{\mathrm{R}}{\rho}_S(t)=&-\int^\infty_0 d\tau [A_3,A_4(-\tau)\rho_S(t)]C_{34}(\tau)\nonumber\\&+[\rho_S(t)A_3(-\tau),A_4]C_{34}(-\tau)\nonumber\\&+[A_4,A_3(-\tau)\rho_S(t)]C_{43}(\tau)\nonumber\\&+ [\rho_S(t)A_4(-\tau),A_3]C_{43}(-\tau),
\end{align}
where
\begin{align}
A_3(t)&=\ket{G}\bra{R} e^{i \epsilon_{\rm R} t}+\ket{L}\bra{D} e^{i (\epsilon_{\rm R} +U) t},\\
A_4(t)&=\ket{R}\bra{G} e^{-i \epsilon_{\rm R} t} + \ket{D}\bra{L}  e^{-i (\epsilon_{\rm R} +U) t},\\
C_{34}(t)&=\mathrm{tr}_{E}(e^{iH_{E_{\rm R}}t}B_{3}e^{-iH_{E_{\rm R}}t}B_4 \rho_{\rm R})\nonumber \\&=\sum_{k_{\rm R}}|t_{k_{\rm R}}|^2 f(\epsilon_{k_{\rm R}}-\mu_{\rm R})e^{i \epsilon_{k_{\rm R}}t},\\
C_{43}(t)&=\mathrm{tr}_{E}(e^{iH_{E_{\rm R}}t}B_{4}e^{-iH_{E_{\rm R}}t}B_3 \rho_{\rm R})\nonumber \\&=\sum_{k_{\rm R}}|t_{k_{\rm R}}|^2 (1-f(\epsilon_{k_{\rm R}}-\mu_{\rm R}))e^{-i \epsilon_{k_{\rm R}}t},
\end{align}
with $\rho_{\rm R}={e^{-\beta_{\rm R} (H_{E_{\rm R}}-\mu_{\rm R}N_{E_{\rm R}})}}/{\mathrm{tr}[e^{-\beta_{\rm R} (H_{E_{\rm R}}-\mu_{\rm R}N_{E_{\rm R}})}]}$.

For the phonon contribution we have $H_{I_{\rm ph}}=A_{5}  B_{5}$ where $A_{5}=\ket{L}\bra{R}+\ket{R}\bra{L}$, $B_{5}=\sum_q h_q(a_q ^\dagger+ a_q)$, and the final term in Eq.~(\ref{HIph}) can be neglected in the weak-coupling limit as it cancels with an energetic shift obtained within the BM approximations~\footnote{See for example Appendix B of Ref.~\cite{IlesSmith14}.}. 
%~\cite{Weiss12,Breuer:2002aa}. 
The resulting phonon Liouvillian reads 
\begin{align}\label{PhononMEPAT}
\mathcal{L}_{\rm ph}{\rho}_S(t)=&-\int^\infty_0 d\tau [A_5,A_5(-\tau)\rho_S(t)]C_{5}(\tau)\nonumber\\&+[\rho_S(t)A_5(-\tau),A_5]C_{5}(-\tau).
\end{align}
Here $A_5(t)=\ket{L}\bra{R} e^{i \Delta t}+\ket{R}\bra{L} e^{-i \Delta t}$  with inter-site energy difference $\Delta=\epsilon_{\rm R}-\epsilon_{\rm L}$. The phonon correlation function $C_{5}(t)=\mathrm{tr}_{E}(B_5(t)B_5 \rho_{\rm ph})=\sum_{q}h_{q}^2 (n(\omega) e^{i \omega t} +(1+n(\omega)) e^{-i \omega t})$ is written in terms of the Bose-Einstein distribution $n(\omega)=(e^{\beta_{\rm ph} \omega}-1)^{-1}$ for a thermal equilibrium phonon bath  $\rho_{\rm ph}={e^{-\beta_{\rm ph} H_{E_{\rm ph}}}}/{\mathrm{tr}(e^{-\beta_{\rm ph} H_{E_{\rm ph}}})}$. 

Combining Eqs.~(\ref{LeftLeadMEPAT}), (\ref{RightLeadMEPAT}) and (\ref{PhononMEPAT}) we arrive at the final WCME for our system evolution, written in the Schr\"odinger picture as
\begin{equation}
\dot{\rho}_S(t)=-i[{H_S,\rho_S(t)}]+(\mathcal{L}_{\rm L}+\mathcal{L}_{\rm R}+\mathcal{L}_{\rm ph})\rho_S(t).
\end{equation}
Throughout the following %our work 
we consider a phonon spectral density 
of Drude-Lorentz form %given as
\begin{equation}
J(\omega)=\sum_qh_q^2\delta(\omega-\omega_q)=\frac{1}{\pi}\frac{2 \lambda \omega \omega_0^2 \gamma}{(\omega_0^2-\omega^2)^2+\gamma^2\omega^2},
\end{equation}
with peak around $\omega_0$, broadening parameter $\gamma$, and reorganisation energy $\lambda$. For both leads we take the wideband limit, $\Gamma_{\rm i}=2\pi\sum_{k_{\rm i}}|t_{k_{\rm i}}|^2\delta(\epsilon-\epsilon_{k_{\rm i}})$.

\subsection{RC mapping for phonon-assisted tunnelling}\label{RCME}
\subsubsection{Non-additive RCME}

\begin{figure}
\center
\includegraphics[width=0.75\linewidth]{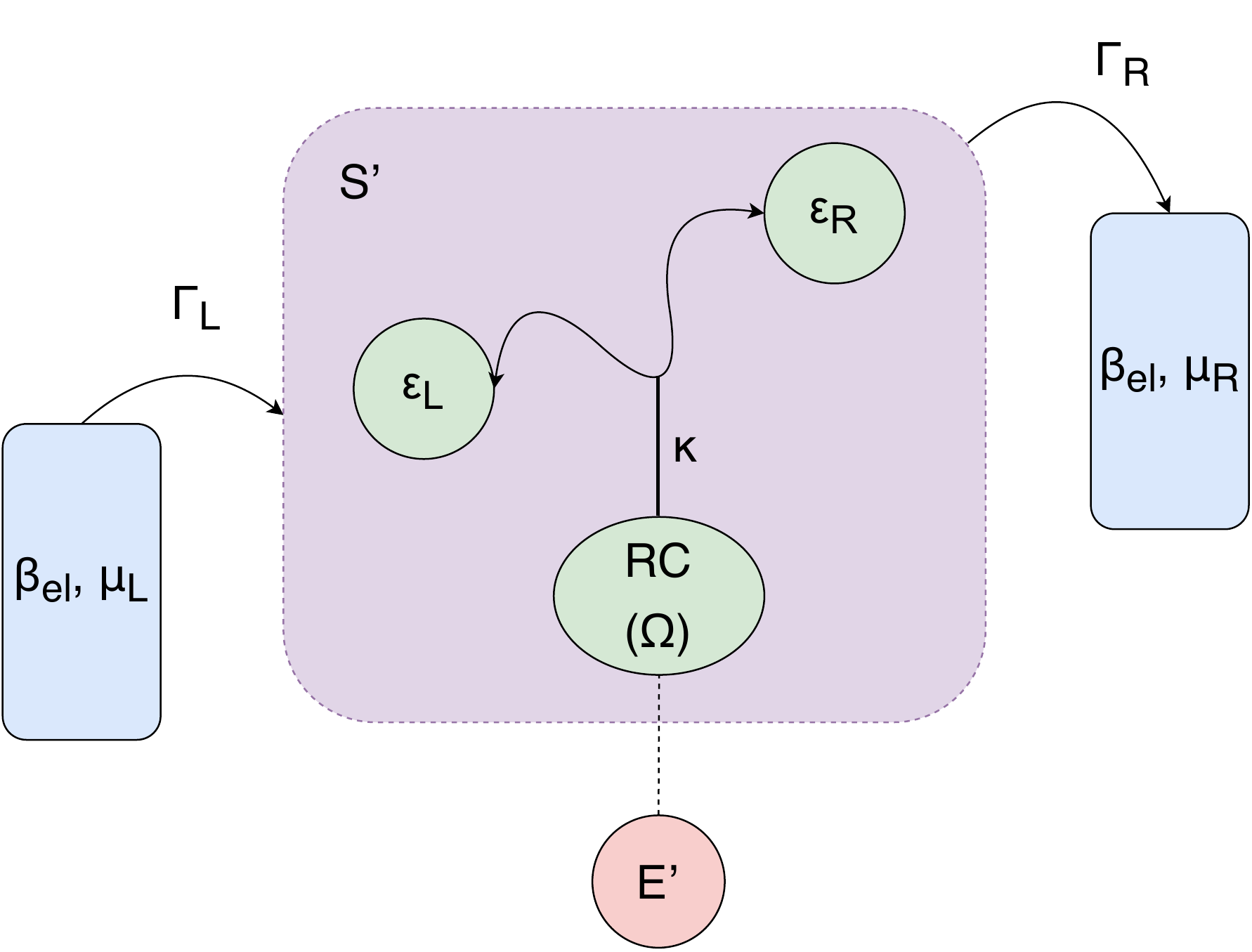}
\caption{Schematic of the reaction coordinate (RC) mapping applied to the two-site %with PAT in a 
thermoelectric model.  
The interactions are treated non-additively, meaning that the leads couple to the augmented system S$'$ which is obtained after the mapping. This is comprised of the original system S and the RC of frequency $\Omega$, interacting with strength $\kappa$. E$'$ represents the residual bosonic environment and other symbols are as defined in Fig.~\ref{fig:PATDiagram}.}
\label{RCDiagramPAT}
\end{figure}

To move beyond the limitations of the WCME treatment of the phonons we now apply the unitary reaction coordinate mapping to include a collective %phonon %environmental 
mode of the phonon environment within an enlarged augmented  system~\cite{IlesSmith14,IlesSmith16,Maguire18,Strasberg16,McConnell19,RCReview} (see Fig.~\ref{RCDiagramPAT}).
Following the %same 
derivations given in \cite{IlesSmith14,Strasberg16,RCReview} and using Eqs.~(\ref{Hsys}) and (\ref{HIph}) the augmented system Hamiltonian becomes %, including the renormalisation term is given in the form
\begin{align}\label{RCSystemHamPAT}
H_{S'}&=(\epsilon_{\rm L} +\lambda) \ket{L}\bra{L} + (\epsilon_{\rm R}+\lambda) \ket{R}\bra{R}\nonumber\\
&+(\epsilon_{\rm L}+\epsilon_{\rm R}+U)\ket{D}\bra{D}\nonumber \\
&+ \kappa (\ket{L}\bra{R}+\ket{R}\bra{L})(a^\dagger + a) + \Omega a^\dagger a.
\end{align}
Here, the RC mode is defined such that
\begin{equation}
\kappa(a^\dagger + a)= \sum_q h_q(a_q^\dagger + a_q)
\end{equation}
with system-RC coupling $\kappa^2=(1/\Omega)\sum_q \omega_qh_q^2$ and mode frequency $\Omega=\omega_0$. 
Note that we must now explicitly include the final term in Eq.~(\ref{HIph}) within our mapped Hamiltonian as the phonons are no longer treated perturbatively, leading in Eq.~(\ref{RCSystemHamPAT}) to the shift $\lambda=\int_0^{\infty}d\omega J(\omega)/\omega$ as the continuum limit of $\sum_q h_q^2/\omega_q$. 

As a result of the mapping the remaining bosonic modes are transformed such that they are now decoupled from the original system, but couple directly to the RC via 
\begin{align}\label{RCInt}
H_{I'_{\rm ph}}&= (a^\dagger + a)\sum_k f_k (b_k^\dagger+ b_k) + (a^\dagger + a)^2 \sum_k \frac{f_k^2}{\nu_k},
\end{align}
where $b_k (b_k^{\dagger})$ are bosonic annihilation (creation) operators. 
To compute the augmented system dynamics, it is not necessary to derive precise forms for the interaction strengths $f_k$ or frequencies $\nu_k$, but rather it suffices to obtain a mapped spectral density $J_{\rm RC}(\nu)=\sum_kf_k^2\delta(\nu-\nu_k)$. This can be achieved through matching Heisenberg equations of motion for the dynamics before and after the mapping, leading to $J_{\rm RC}(\nu)=\gamma\nu/2\pi\omega_0$~\cite{IlesSmith14,Strasberg16,RCReview}. As the mapping has no effect on the fermionic leads, the total Hamiltonian now becomes
\begin{equation}
H=H_{S'}+H_{E_{\rm L}}+H_{E_{\rm R}}+H_{E'_{\rm ph}}+H_{I_{\rm L}}+H_{I_{\rm R}}+H_{I'_ {\rm ph}},
\end{equation} 
with $H_{E'_{\rm ph}}=\sum_k\nu_kb_k^{\dagger}b_k$.

Following the %same 
procedure described in~\cite{McConnell19} we trace out the leads and the {\it residual} bosonic bath within the BM approximations to derive a master equation for the augmented system density operator $\rho_{S'}(t)$, which describes the original system plus RC, and is given by
\begin{align}\label{fullmePAT}
\dot{\rho}_{S'}(t)&=-i[H_{S'},\rho_{S'}(t)]-[A_{\rm ph},[\chi_{\rm ph},\rho_{S'}(t)]]\nonumber\\&+[A_{\rm ph},\{\phi_{\rm ph},\rho_{S'}(t)\}]
\nonumber\\&-\Gamma_{\rm L} \Big([A_1,\chi_{2,{\rm L}}] \rho_{S'}(t)]+[\rho_{S'}(t)\phi_{2,{\rm L}},A_1]\nonumber\\
&+[A_2,\phi_{1,{\rm L}} \rho_{S'}(t)]+[\rho_{S'}(t)\chi_{1,{\rm L}},A_2]\Big)
\nonumber\\&-\Gamma_{\rm R} \Big([A_3,\chi_{4,{\rm R}} \rho_{S'}(t)]+[\rho_{S'}(t)\phi_{4,{\rm R}},A_3]\nonumber\\&+[A_4,\phi_{3,{\rm R}} \rho_{S'}(t)]+[\rho_{S'}(t)\chi_{3,{\rm R}},A_4]\Big).
\end{align}
Here $A_{m}$ for $m=1,2,3,4$ are the same system operators as defined in the previous section, $A_{\rm ph}=(a^\dagger +a)$ and 
\begin{align}
\chi_{\rm ph}
&= \frac{\pi}{2} \sum_{j k} J_{\rm RC}(\eta_{jk})\mathrm{coth}\Big(\frac{\beta \eta_{jk}}{2}\Big) A_{{\rm ph}_{jk}}\ket{\psi_j}\bra{\psi_k},\label{RCPhononComp1}\\
\phi_{\rm ph}
&= \frac{\pi}{2} \sum_{j k} J_{\rm RC}(\eta_{jk}) A_{{\rm ph}_{jk}}\ket{\psi_j}\bra{\psi_k}\label{RCPhononComp2},\\
\chi_{m,{\rm i}}&=\sum_{jk} A_{m_{j,k}}\ket{\psi_j}\bra{\psi_k}f_{\rm i}(\pm\eta_{jk}),\label{leadnonadd1}\\ 
\phi_{m,{\rm i}}&= \sum_{jk} A_{m_{j,k}}\ket{\psi_j}\bra{\psi_k}(1 - f_{\rm i}(\pm\eta_{jk})),\label{leadnonadd2}
\end{align}
where ${\rm i}={\rm L},{\rm R}$ as before.  
Defining the system-RC eigenbasis through $H_{S'}\ket{\psi_j}=\psi_j\ket{\psi_j}$, we have $A_{m_{jk}}=\bra{\psi_j}A_m\ket{\psi_k}$, $A_{{\rm ph}_{jk}}=\bra{\psi_j}A_{\rm ph}\ket{\psi_k}$ and $\eta_{jk}=\psi_j -\psi_k$. Eqs.~(\ref{RCPhononComp1}) and (\ref{RCPhononComp2}) describe the action of the residual bosonic bath on the mapped system through its coupling to the RC. Eqs.~(\ref{leadnonadd1}) and (\ref{leadnonadd2}) encode the influence of the leads, and have been derived within the full system-RC eigenbasis so that the leads are ``aware" of the strong system-phonon coupling. Thus the treatment is inherently non-additive with respect to the lead and phonon couplings, in contrast to the WCME. Note that, despite the BM approximation applied to the residual bosonic bath, tracing over the RC in $\rho_{S'}(t)$ results in non-Markovian dynamics for the reduced system state $\rho_S(t)$ that is also valid at large phonon reorganisation energies $\lambda$, again in contrast to the WCME ~\cite{IlesSmith14,IlesSmith16,Strasberg16}.
 
\subsubsection{Additive RCME}

As we have just seen, when phonon coupling is strong the RC treatment lends itself naturally to a non-additive description of the 
fermionic leads in our thermoelectric model. %in our thermoelectric system. 
This may not, however, be the case 
for other techniques that can treat phonons beyond the weak-coupling limit, and in such situations it might only be feasible to include further environments phenomenologically. It is thus important to assess the impact of such a simplifying assumption on thermoelectric performance, which can be done through an additive reaction-coordinate master equation (aRCME).   

The aRCME is obtained straightforwardly from the full RCME of Eq.~(\ref{fullmePAT}) by replacing the left and right lead terms with their respective weak-coupling counterparts given in Eqs.~(\ref{LeftLeadMEPAT}) and~(\ref{RightLeadMEPAT}). Specifically, the aRCME reads %master equation we have for the PAT set-up is
\begin{align}\label{addmePAT}
\dot\rho_{\rm S'}(t)&=-i[H_{\rm S'},\rho_{\rm S'}(t)]+(\mathcal{L}_{\mathrm{L}}\otimes\mathbb{I}_{\rm RC})\rho_{\rm S'}(t)\nonumber\\&+(\mathcal{L}_{\mathrm{R}}\otimes\mathbb{I}_{\rm RC})\rho_{\rm S'}(t)-[A_{\rm ph},[\chi_{\rm ph},\rho_{\rm S'}(t)]]\nonumber\\&+[A_{\rm ph},\{\phi_{\rm ph},\rho_{\rm S'}(t)\}],
\end{align}
where $\mathbb{I}_{\rm RC}$ denotes the identity operator applied within the RC Hilbert space, such that the leads couple only to the original (unmapped) two-site system and are unaffected by the strong phonon coupling.

\subsection{Counting Statistics}

We characterise the thermoelectric performance of our setup by tracking the counting statistics %number 
of electrons that enter or leave the right lead under steady-state conditions. Following Refs.~\cite{Flindt05,Flindt10} we define the matter current flow (first cumulant or mean $\langle\langle I\rangle\rangle$) and the associated noise (second cumulant or variance $\langle\langle I^2\rangle\rangle$) %within our system 
as
\begin{align}\label{current}
\langle\langle I\rangle\rangle&=\langle\langle\tilde{0}|\mathcal{I}|0\rangle\rangle,\\
  \langle\langle I^2 \rangle\rangle &= \langle\bra{\tilde{0}} \mathcal{J}\ket{0}\rangle -2\langle\bra{\tilde{0}} \mathcal{IRI}\ket{0}\rangle,\label{NoiseLiouv}
\end{align}
where $\mathcal{I}=\mathcal{I}^+-\mathcal{I}^-$, $\mathcal{J}=\mathcal{I}^++\mathcal{I}^-$, $|0\rangle\rangle $ is the right eigenvector corresponding to the zero eigenvalue of the Liouvillian under consideration (WCME, RCME, or aRCME), and $\langle\langle\tilde{0}|$ is the identity operator. %or corresponding left eigenvector of the zero eigenvalue. 
The current operator $\mathcal{I}^+$ ($\mathcal{I}^-$) represents the terms in the relevant Liouvillian that add (remove) electrons to (from) the right lead. 
Defining projection operators $\mathcal{P}=\ket{0}\rangle\langle\bra{\tilde{0}}$ and $\mathcal{Q}=\mathbb{I}-\mathcal{P}$ 
we have a pseudo-inverse $\mathcal{R}=\mathcal{QL}^{-1}\mathcal{Q}$, %is also presented, and 
which is well defined as the inversion is only performed on the subspace spanned by $\mathcal{Q}$.

\begin{figure*}[t]
     \subfloat[][$\beta_{\rm L}=0.1 \beta_{\rm R}$, $V\beta_{\rm R} =0.1$]{{\includegraphics[width=0.45\textwidth]{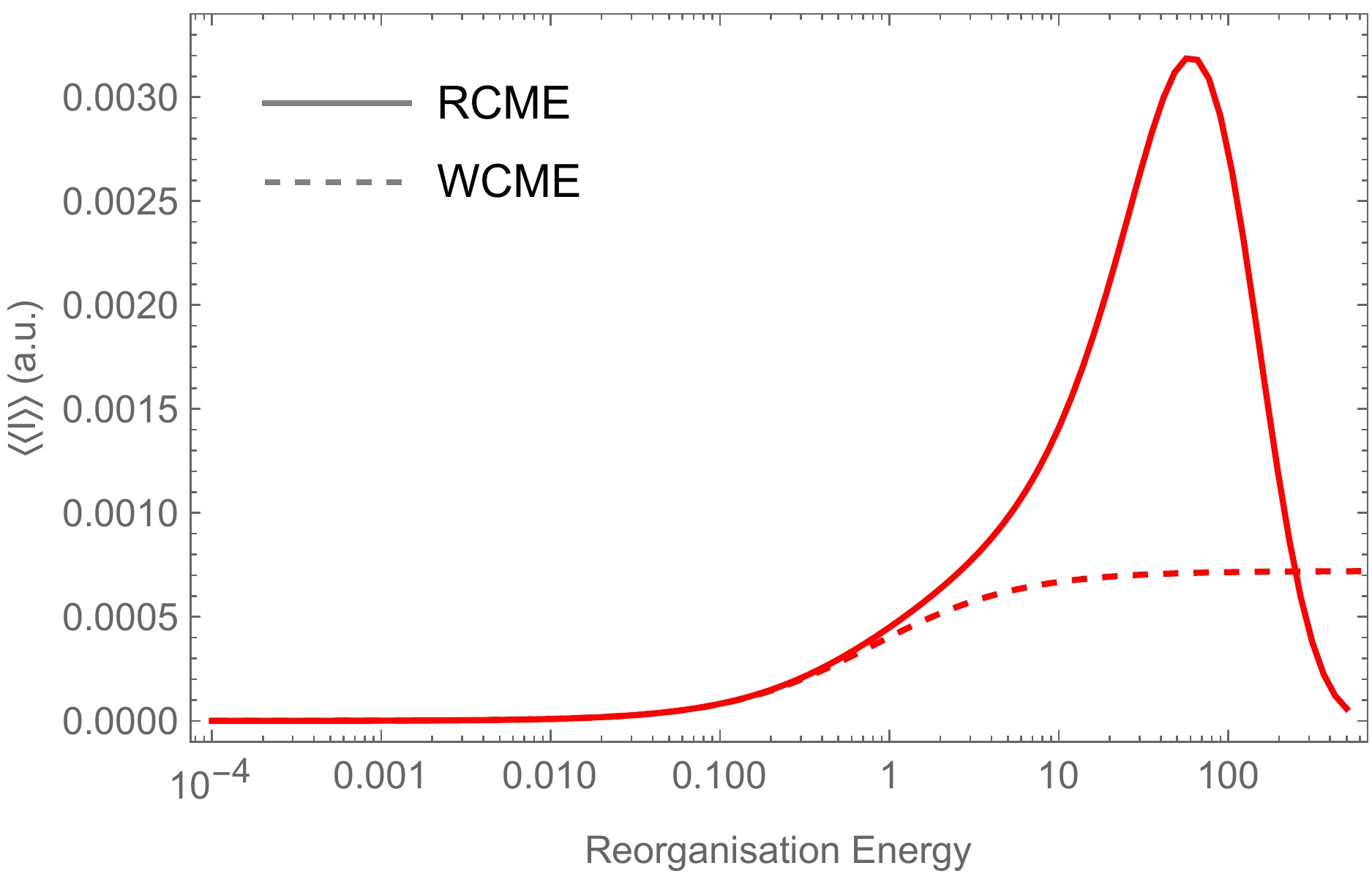}}}%
     \qquad    
    \subfloat[ ][$\lambda\beta_{\rm R} =3$, $V \beta_{\rm R} =0.1$]{{\includegraphics[width=0.45\textwidth]{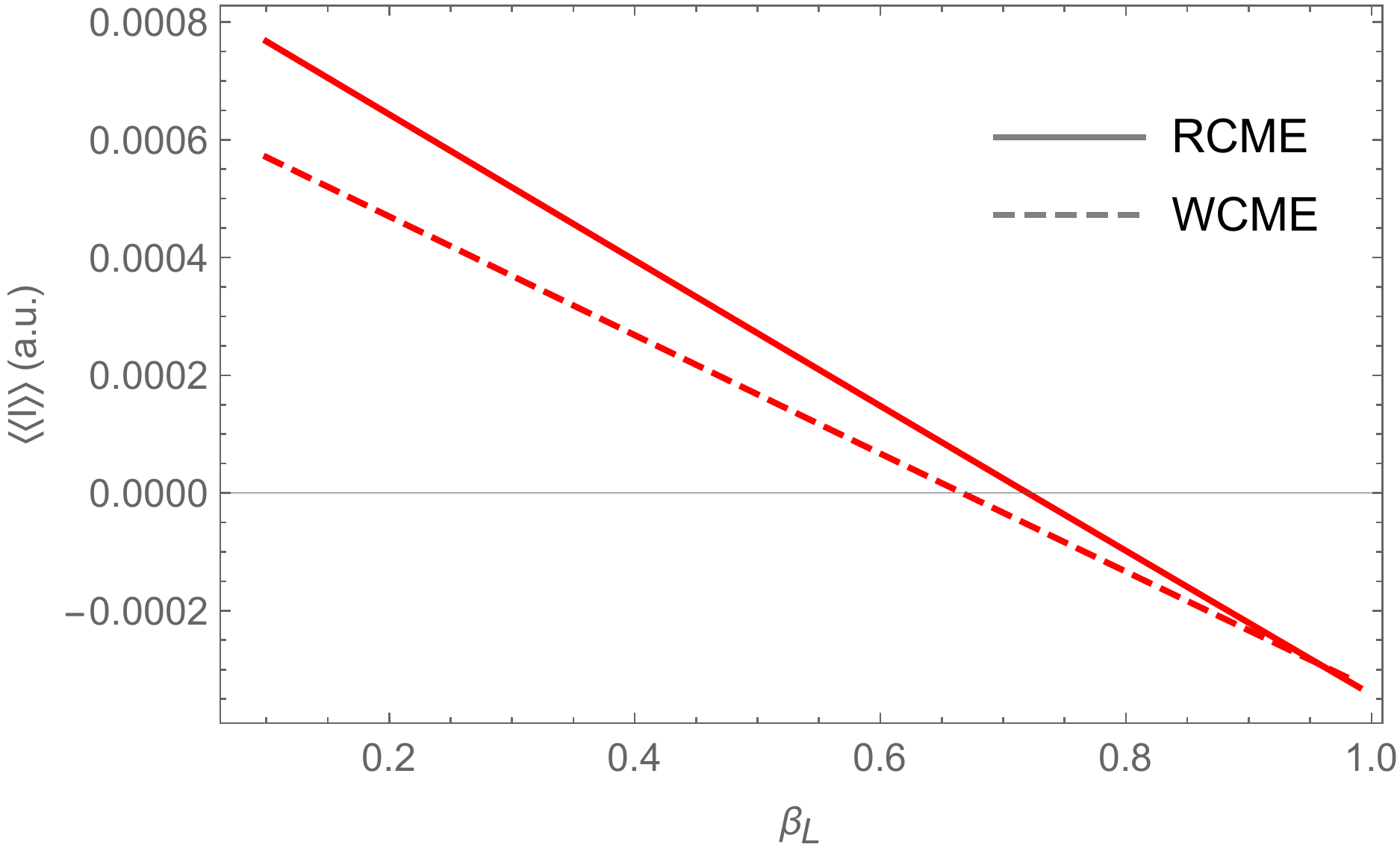}}}%
     \qquad \\ 
    \subfloat[ ][$\lambda\beta_{\rm R} =3$, $\beta_{\rm L}=0.1 \beta_{\rm R}$]{{\includegraphics[width=0.45\textwidth]{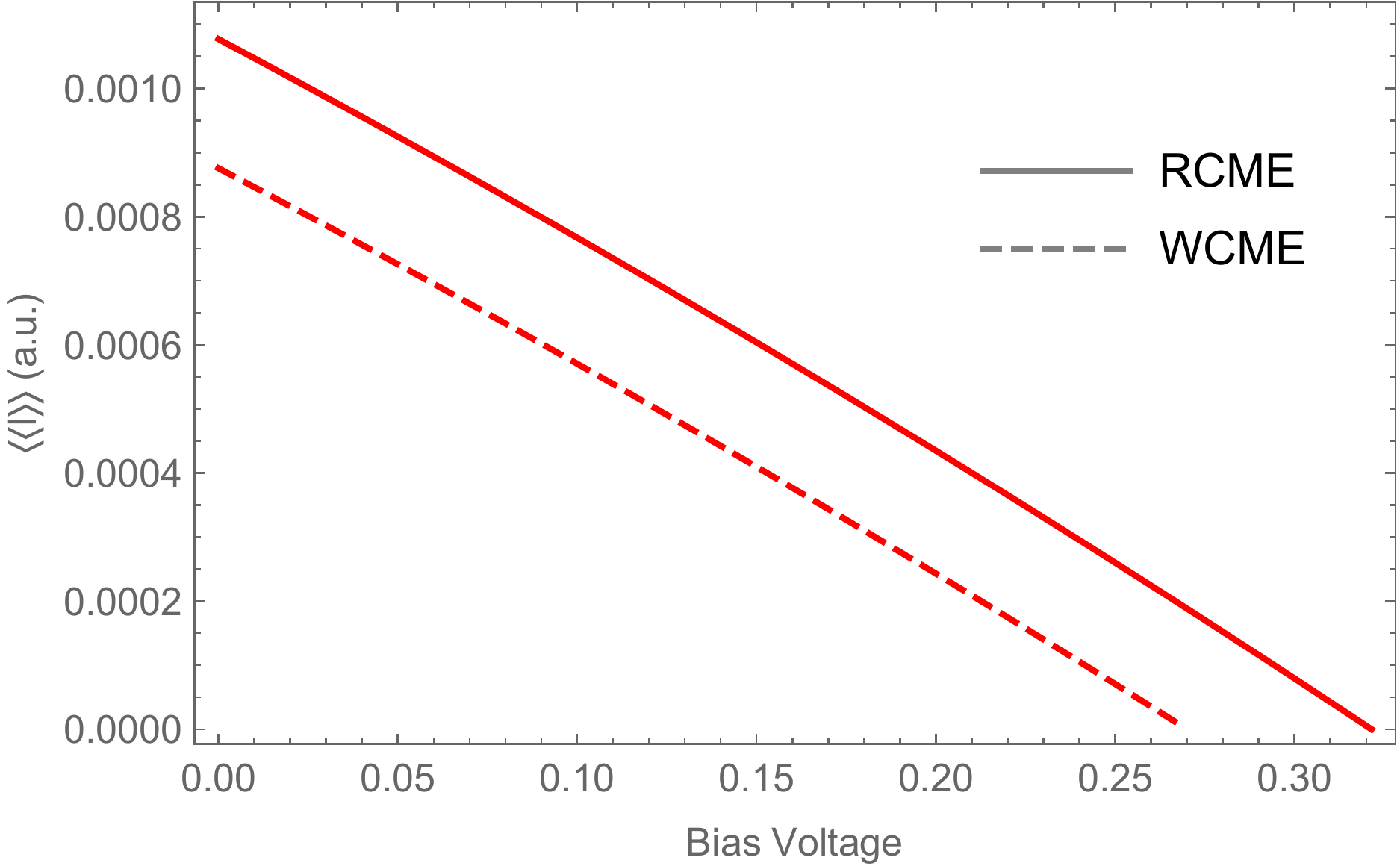}}}%
     \qquad 
     \subfloat[][$\lambda\beta_{\rm R} =3$, $\beta_{\rm L}=0.1 \beta_{\rm R}$]{{\includegraphics[width=0.45\textwidth]{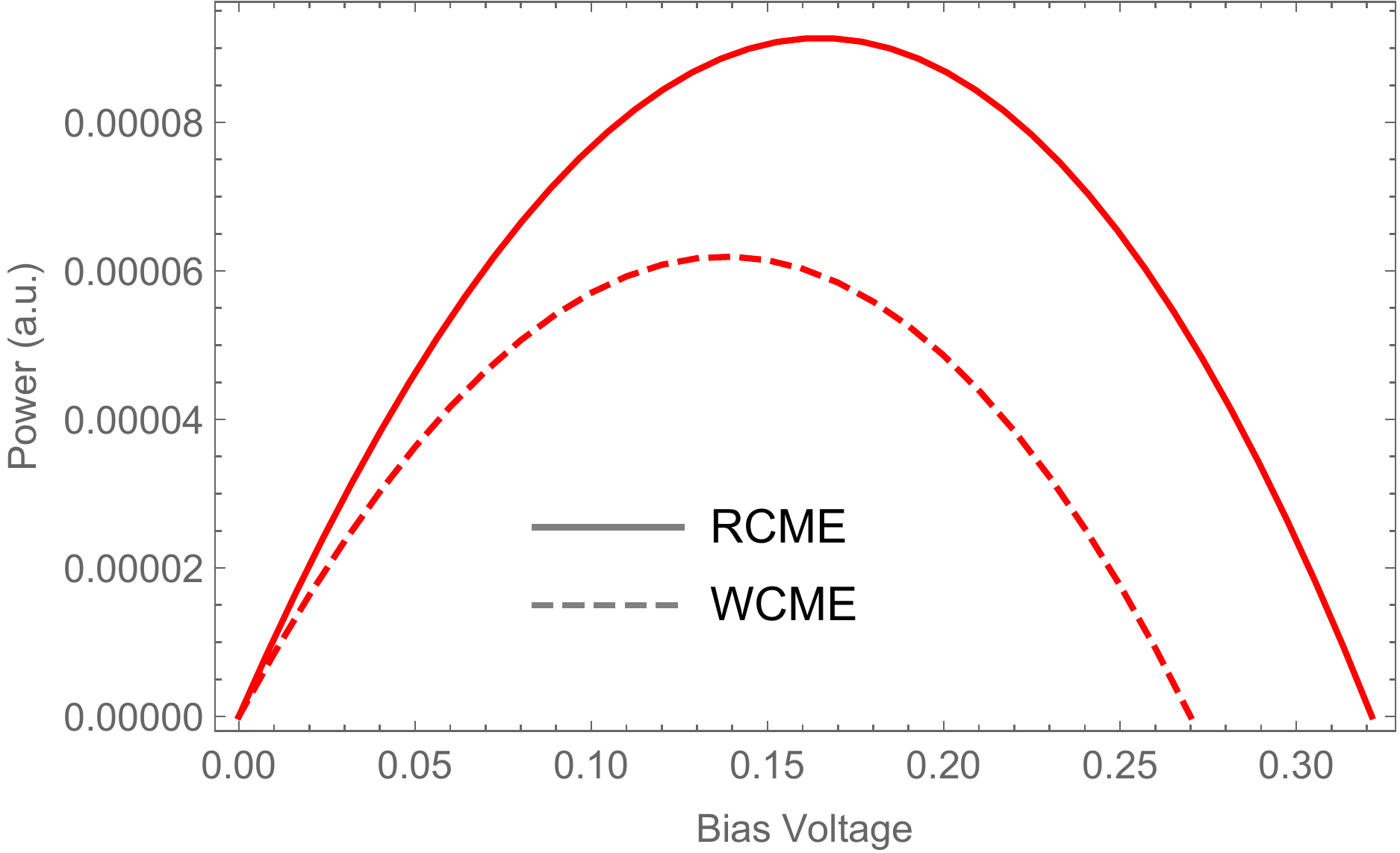}}}
     \caption{Current as a function of (a) reorganisation energy $\lambda$ (in units of $\beta_{\rm R}^{-1}$), (b) inverse temperature of the left lead $\beta_{\rm L}$ (in units of $\beta_{\rm R}$), and (c) bias voltage $V$ (in units of $\beta_{\rm R}^{-1}$). Plot (d) gives the power output as a function of bias voltage for fixed reorganisation energy and temperature gradient. Other parameters used are $\Delta\beta_{\rm R}=2, \omega_0\beta_{\rm R}=\gamma\beta_{\rm R}=100$ and $\Gamma_{\rm L}\beta_{\rm R}=\Gamma_{\rm R}\beta_{\rm R}=0.1$.}
      \label{fig:PATThermoelectricLeadResourceCurrent}
\end{figure*}

Obtaining the thermoelectric power output $P$ is straightforward once we have the current, %now that we have the current, as
\begin{equation}
P= (\mu_{\rm R}-\mu_{\rm L})\langle\langle I \rangle\rangle.
\end{equation}
The thermoelectric efficiency is defined as
\begin{equation}
\eta=\frac{P}{Q_{\mathrm{in}}},
\end{equation} 
where %$\eta$ is the thermoelectric efficiency and 
$Q_{\mathrm{in}}$ is the heat flow entering the system. 
In Sec.~\ref{leftresource} we have $T_{\rm L}> T_{\rm R}=T_{\rm ph}$ such that the left lead acts as the resource and $Q_{\mathrm{in}}=Q_{\rm L}$. Since the total number of electrons is conserved, the steady-state current from the left lead is given by $\langle\langle I\rangle\rangle$. In the following examples we always take $U$ to be by far the largest energy scale, such that we work in the Coulomb blockade regime and the state $\ket{D}$ is never occupied. In the WCME and the aRCME this means that each electron has an associated energy $\epsilon_{\rm L}$, and the left lead energy current becomes $I_{\rm L}^E=\epsilon_{\rm L}\langle\langle I\rangle\rangle$. In the full RCME however, the left lead now couples to a manifold of states within the augmented system and we cannot simply consider a single energy scale $\epsilon_{\rm L}$. Instead, we use the fact that in the steady-state ${\rm tr}(H_{S'}\dot{\rho}_{S'})=0$ to obtain the left lead energy current from Eq.~(\ref{fullmePAT}) as
\begin{align}
I_L^E=-\Gamma_{\rm L}{\rm tr}& \Big(H_{S'}\big\{[A_1,\chi_{2,{\rm L}}] \rho_{S'}(\infty)]+[\rho_{S'}(\infty)\phi_{2,{\rm L}},A_1]\nonumber\\
&+[A_2,\phi_{1,{\rm L}} \rho_{S'}(\infty)]+[\rho_{S'}(\infty)\chi_{1,{\rm L}},A_2]\big\}\Big).
\end{align}
Here $\rho_{S'}(\infty)$ is the RCME steady-state obtained from Eq.~(\ref{fullmePAT}) by setting $\dot{\rho}_{S'}=0$. Note that we could also have calculated $I_L^E$ for the WCME and the aRCME in an analogous way, though the result would be the same as setting $I_{\rm L}^E=\epsilon_{\rm L}\langle\langle I\rangle\rangle$ for these cases. The heat flow can be calculated in all cases from $Q_{\rm L}=I_{\rm L}^E-\mu_{\rm L}\langle\langle I\rangle\rangle$ using the appropriate energy current.

\begin{figure*}[t]
     \subfloat[][$\lambda \beta_{\rm R}=3$, $\beta_{\rm L} =0.1 \beta_{\rm R}$]{{\includegraphics[width=0.45\textwidth]{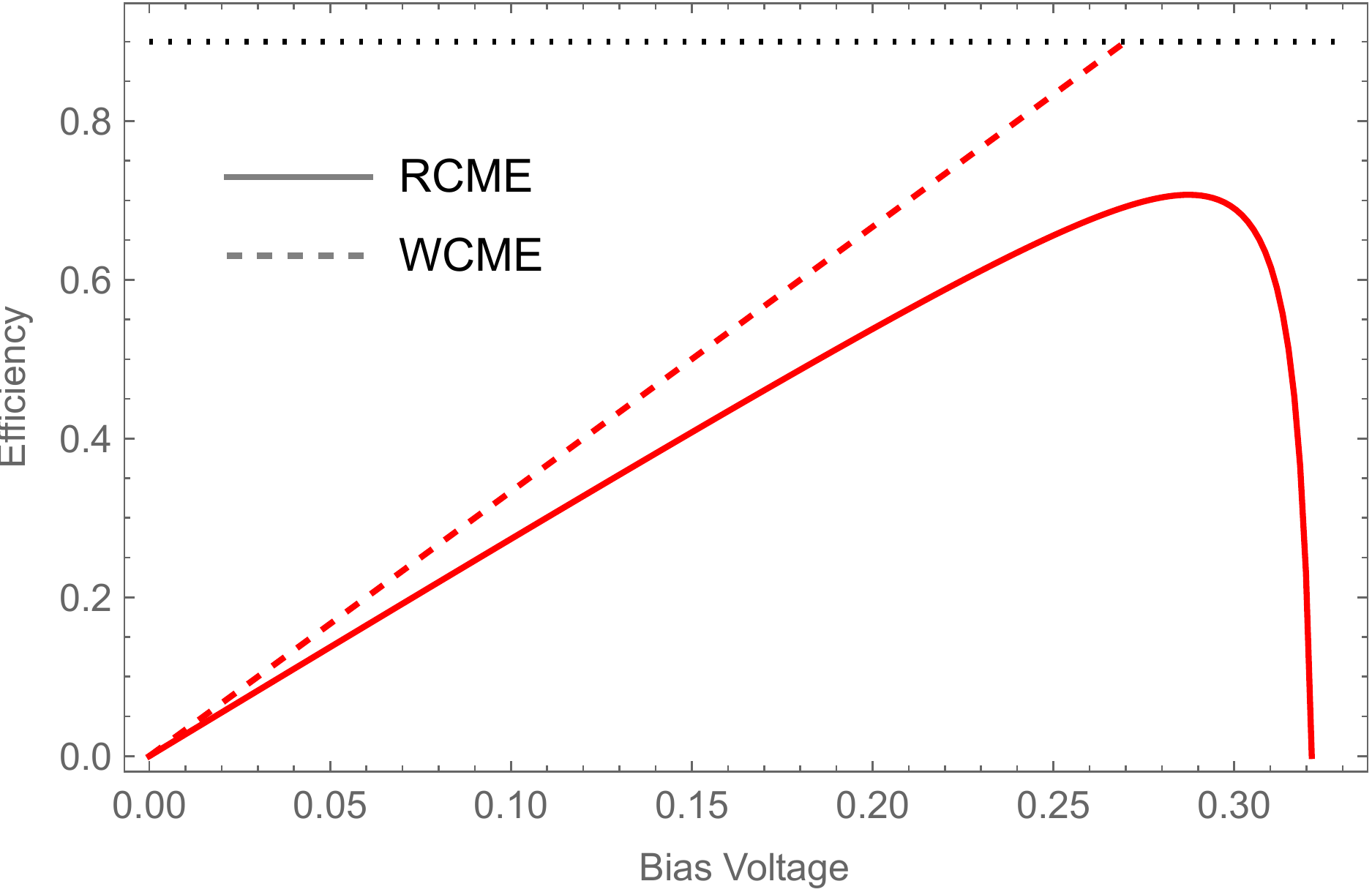}}}%
     \qquad    
    \subfloat[ ][$\beta_{\rm L} =0.1 \beta_{\rm R}$, $V \beta_{\rm R}=0.1$]{{\includegraphics[width=0.45\textwidth]{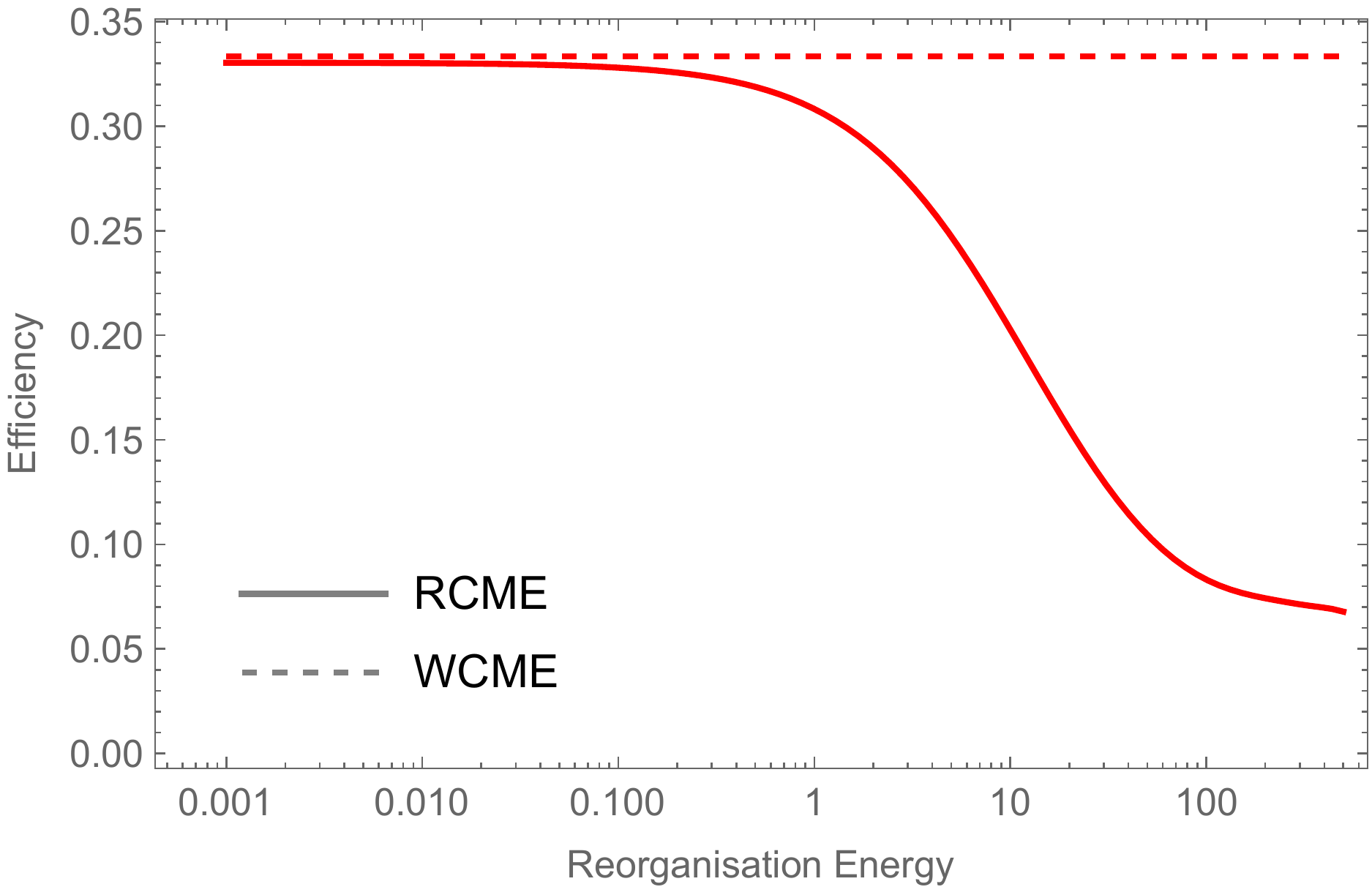}}}%
     \qquad \\ 
    \subfloat[ ][$\beta_{\rm L} =0.1 \beta_{\rm R}$, $V \beta_{\rm R}=0.1$]{{\includegraphics[width=0.45\textwidth]{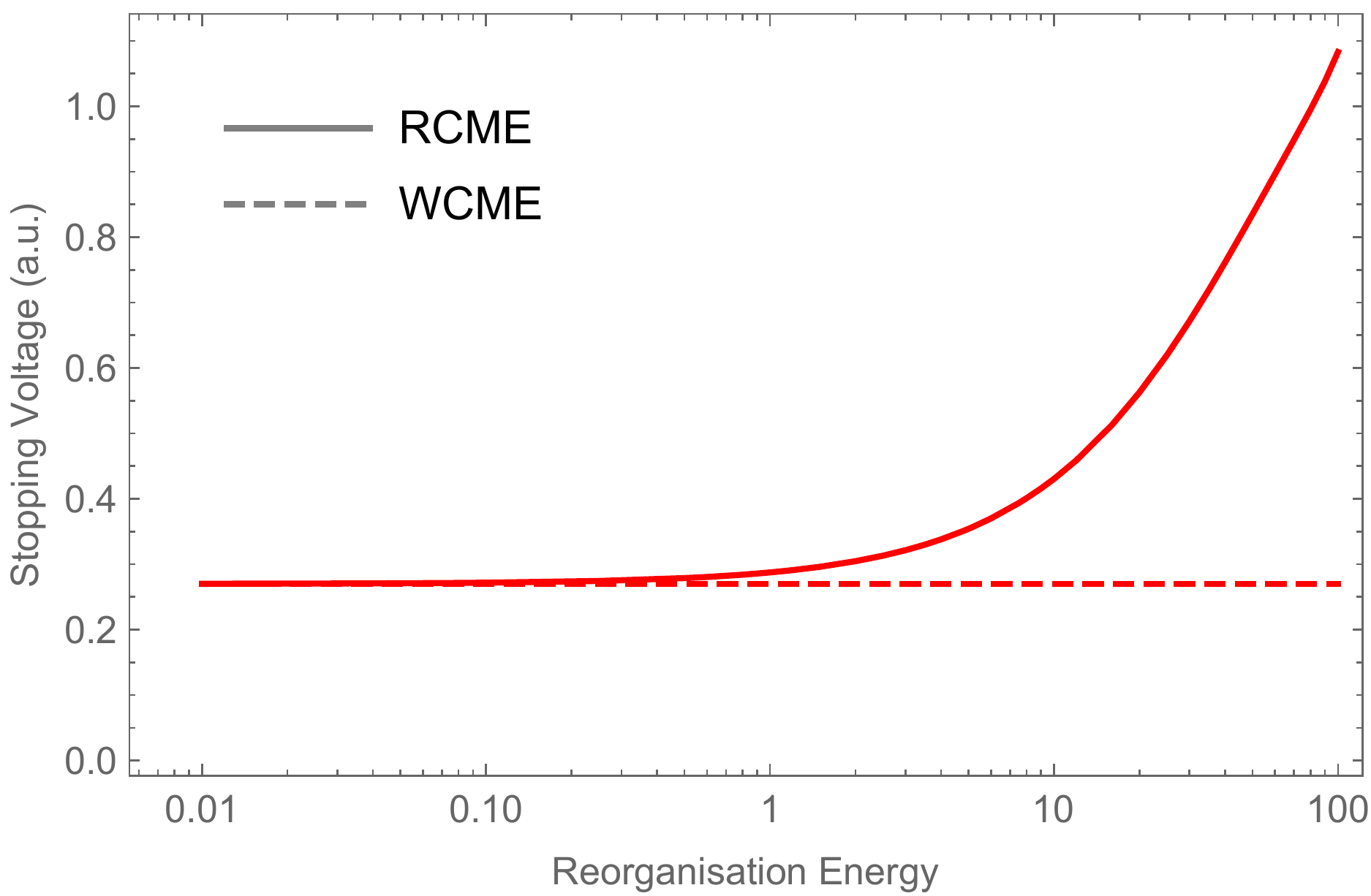}}}%
     \qquad 
     \subfloat[ ][$\beta_{\rm L} =0.1 \beta_{\rm R}$]{{\includegraphics[width=0.45\textwidth]{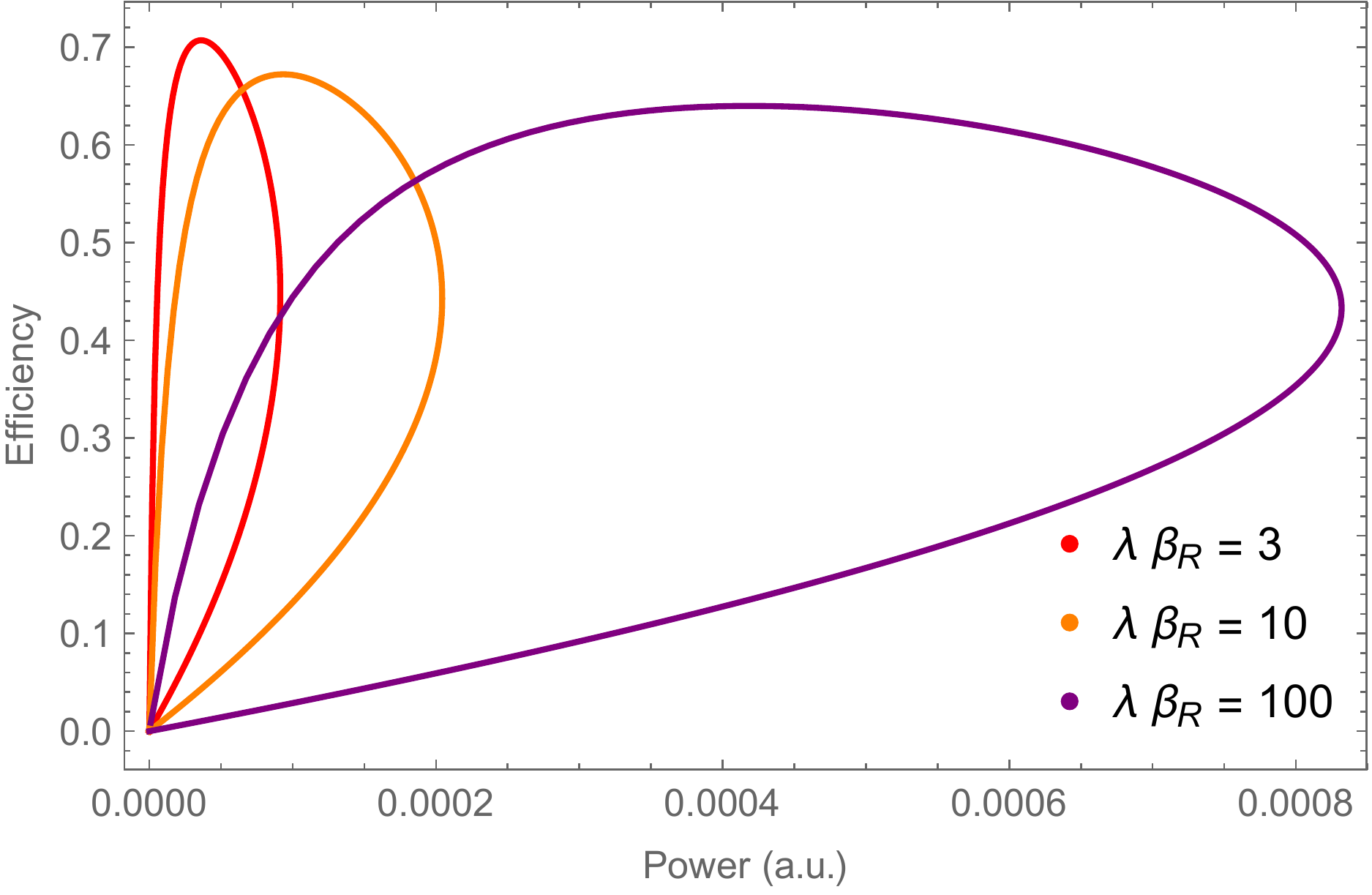}}}
     \caption{Efficiency as a function of (a) bias voltage %for a fixed temperature gradient and reorganisation energy 
     (black dotted line shows the Carnot efficiency) and (b) %as a function of 
     reorganisation energy. %for a fixed temperature gradient and bias voltage. 
     (c) Stopping voltage as a function of reorganisation energy. %for a fixed temperature gradient and bias voltage. 
     (d) Parametric plot of power and efficiency generated by altering the bias voltage for fixed temperature gradient and various %$\lambda$ 
reorganisation energies %$\beta_{\rm L}$ 
within the RCME treatment. Other parameters used are $\Delta\beta_{\rm R}=2, \omega_0\beta_{\rm R}=\gamma\beta_{\rm R}=100$ and $\Gamma_{\rm L}\beta_{\rm R}=\Gamma_{\rm R}\beta_{\rm R}=0.1$.}
      \label{fig:PATThermoelectricLeadResourceEff}
\end{figure*}

In Sec.~\ref{phononresource}, it is the phonons that act as a resource rather than the left lead, such that $Q_{\rm in}=Q_{\rm ph}$. As no work is exchanged between the phonon environment and the system, $I^E_{\rm ph}=Q_{\rm ph}$ and the phonon heat flow is equal to the phonon energy current. We may then use conservation of energy in the steady-state to calculate the heat flow from $I^E_{\rm L}+I^E_{\rm R}+I^E_{\rm ph}=0$, where $I^E_{\rm R}$ here is positive for energy {\it entering} the system from the right lead. 

\section{Thermoelectric Regime I: Left Lead Resource}\label{leftresource}

In our first example %thermoelectric regime we %want to
we consider the impact of strong electron-phonon coupling when a temperature gradient is introduced between the left and right leads. %rather than between the phonon bath and the leads. 
The phonon bath is held at the same temperature as the right lead (the colder of the two) so that it is only able to act as a transport mediator without being a thermodynamic resource. 

We begin in Fig.~\ref{fig:PATThermoelectricLeadResourceCurrent}(a) by studying the thermoelectric current flow as a function of phonon coupling (reorganisation energy, $\lambda$) for a fixed bias voltage. 
We see that at small reorganisation energies the WCME and RCME agree, as expected, and both predict an increase in current with $\lambda$ due to %the increased strength of 
stronger phonon-mediated tunneling between the left and right dot. As the phonon reorganisation energy becomes larger, and we move out of the validity range of the BM approximations, the WCME incorrectly predicts a plateau in the current due to saturation of the available transport channels. Specifically, in the WCME transport proceeds via sequential incoherent transitions with electrons tunneling in turn between the left lead and the left dot, then via phonons between the left and right dots, and finally tunneling onto the right lead. As the reorganisation energy is increased, phonon mediated tunneling between the left and right dots becomes increasingly rapid (the WCME single-phonon rate is proportional to $\lambda$), until the tunneling rate between the dots and the leads becomes the limiting factor (saturation) and the current plateau. 

For small reorganisation energies, transport proceeds in the same manner within the RCME and single-phonon processes dominate. 
In contrast, an enhancement of current is observed in the RCME predictions for $\lambda\beta_{\rm R}\sim1-100$. In this regime multi-phonon processes that are ignored in the WCME,  
but are captured through the coupling of the leads to the numerous eigenstates of the augmented system within the RCME, become increasingly important, %within the RC approach. 
opening up further transport channels that allow for greater current flow. %as more transport channels are utilised. 
Furthermore, the energy levels 
of the dominant transport channels are renormalised by the strong system-phonon coupling; they are increased in comparison to $\mu_L$ but differences between them are reduced in comparison to the WCME, %these states is reduced , 
allowing %the phonon assisted 
tunnelling to occur more readily. 
Nevertheless, for very large reorganisation energies the RCME predicts a complete suppression of current flow due to electron blockade, which again is not captured within the WCME. Similar to Franck-Condon blockade~\cite{Koch05,PhysRevB.74.205438}, in the RCME displacement of the RC becomes large at strong vibrational coupling. Hence overlaps between the augmented system states responsible for transport and the empty state, for which the RC is not displaced, are substantially suppressed, leading in turn to the reduction of current flow.

We now consider variations in the left lead temperature, shown in Fig.~\ref{fig:PATThermoelectricLeadResourceCurrent}(b) for 
an intermediate reorganisation energy of $\lambda \beta_{\rm R} =3$, where %we expect 
the RCME predicts a moderate increase in current compared to the WCME. Both the WCME and RCME predict current flow decreases as $\beta_{\rm L}$ increases 
due to the %smaller 
reduction in temperature gradient between the leads, which suppresses the thermoelectric activity. 
Both also predict a crossover to negative current flow %as $\beta_{\rm L}$ becomes larger, in both cases we have 
once the relevant stopping voltage %, $V_S$, 
is surpassed %. Then 
and the current then 
follows the chemical potential (rather than temperature) gradient. In fact, the RCME predicts a slightly more resilient thermoelectric, with the chosen value of $V$ acting as the stopping voltage for a smaller temperature gradient than for the WCME. %\par

This can also be seen in Fig.~\ref{fig:PATThermoelectricLeadResourceCurrent}(c), where the RCME predicts a larger current for all bias voltages and  
an extended stopping voltage compared to the WCME prediction of $V_S =(\epsilon_{\rm L} -\mu_{\rm L})(\beta_{\rm R}-\beta_{\rm L})/\beta_{\rm R}$. Again, this increase in stopping voltage is due to the non-additive interplay between the phonon bath and the fermionic leads, encapsulated through the RC mapping and subsequent master equation derivation, which effectively increases the relevant system energies in comparison to $\mu_{\rm L}$. %For the same reason, 
We see a similar behaviour in the thermoelectric power in Fig.~\ref{fig:PATThermoelectricLeadResourceCurrent}(d). For a constant bias, the power as a function of reorganisation energy is of the same form as the current in Fig. \ref{fig:PATThermoelectricLeadResourceCurrent}(a), simply scaled by the value of $V$.

Having considered the current flow and power output we now turn our attention to the associated thermoelectric efficiency, shown as a function of bias voltage in Fig.~\ref{fig:PATThermoelectricLeadResourceEff}(a).  
In the WCME case the efficiency increases linearly with $V$ %the bias voltage, %running from zero efficiency at zero bias 
up to a maximum of the Carnot efficiency $\eta_C$ at the stopping voltage. The RCME prediction is qualitatively different, however, and we see that the increase in power due to strong phonon coupling comes at the expense of a reduced efficiency for heat to work conversion, and the inability of the system to reach the Carnot efficiency even for vanishing power output. 
This drop in efficiency is caused by the power decreasing towards zero more quickly than the heat flow, with the left lead providing heat energy that is preferentially absorbed by the phonon bath rather than promoting electrons to tunnel against the increasing bias. 
This can be thought of as a %drop in efficiency is a 
cost imposed by the strong electron-phonon coupling, and the resultant power-efficiency trade off can be seen %clearly from comparison of 
by comparing the RCME curves in Fig.~\ref{fig:PATThermoelectricLeadResourceEff}(b), which shows efficiency as a function of phonon coupling strength, and in Fig.~\ref{fig:PATThermoelectricLeadResourceCurrent}(a). 
In contrast, the WCME efficiency is unaffected by increasing the phonon reorganisation energy, another failing of the BM approximations. %In the weak coupling case the efficiency can be simplified, giving
Instead, we have 
\begin{equation}
\eta^{\mathrm{WCME}}=\frac{P}{Q_{in}}=\frac{\langle\langle I\rangle\rangle V}{(\epsilon_{\rm L}-\mu_{\rm L})\langle\langle I\rangle\rangle}=\frac{V}{\epsilon_{\rm L}-\mu_{\rm L}},
\end{equation}
which is independent of $\lambda$, and for $V \rightarrow V_S$ becomes $\eta^{\mathrm{WCME}}=(\beta_{\rm R}-\beta_{\rm L})/\beta_{\rm R}=\eta_C$. 
Likewise, the WCME fails to predict any variation of the stopping voltage with phonon coupling, as shown in Fig.~\ref{fig:PATThermoelectricLeadResourceEff}(c). On the other hand, the RCME predicts an increase in stopping voltage with $\lambda$, consistent again with a larger parameter regime supporting thermoelectric behaviour at strong phonon coupling.

\begin{figure}[t]
     \subfloat[][$\beta_{\rm L} =0.1 \beta_{\rm R}$, $V \beta_{\rm R}=0.1$]{{\includegraphics[width=0.45\textwidth]{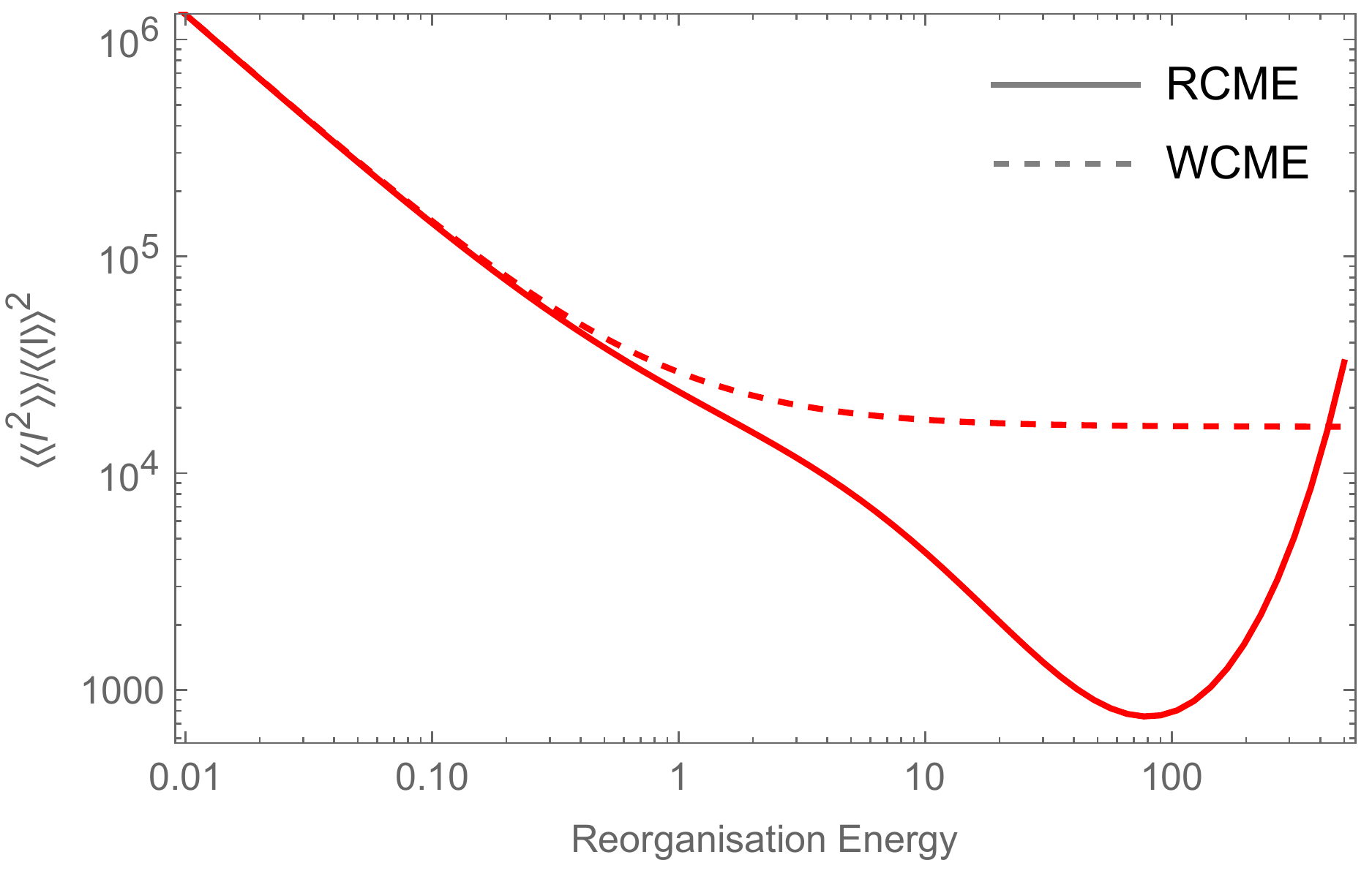}}}%
    \qquad    
 \subfloat[][$\beta_{\rm L} =0.1 \beta_{\rm R}$, $\lambda \beta_{\rm R}=3$]{{\includegraphics[width=0.45\textwidth]{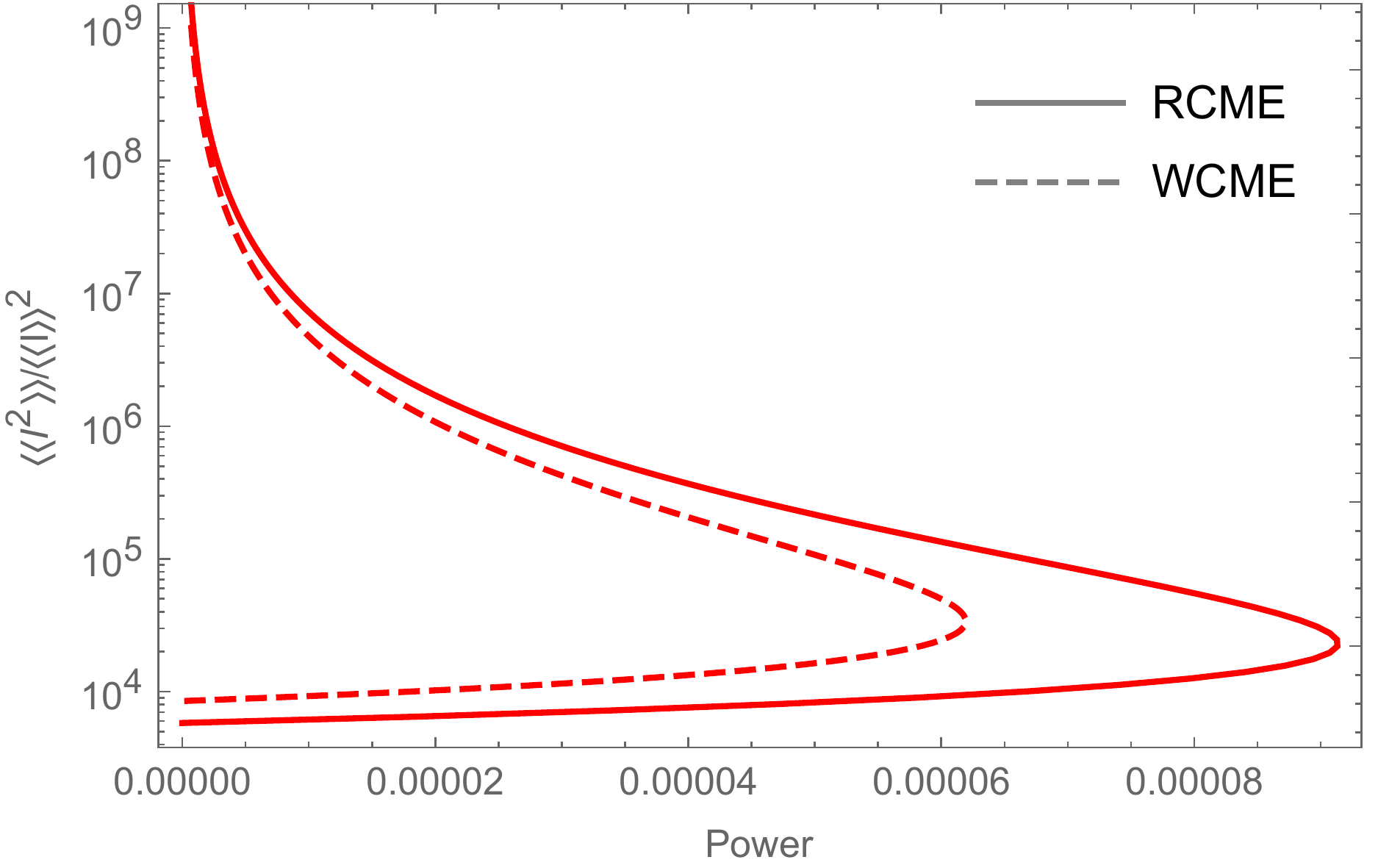}}}
     \qquad
 \subfloat[][$\beta_{\rm L} =0.1 \beta_{\rm R}$, $\lambda \beta_{\rm R}=3$]{{\includegraphics[width=0.45\textwidth]{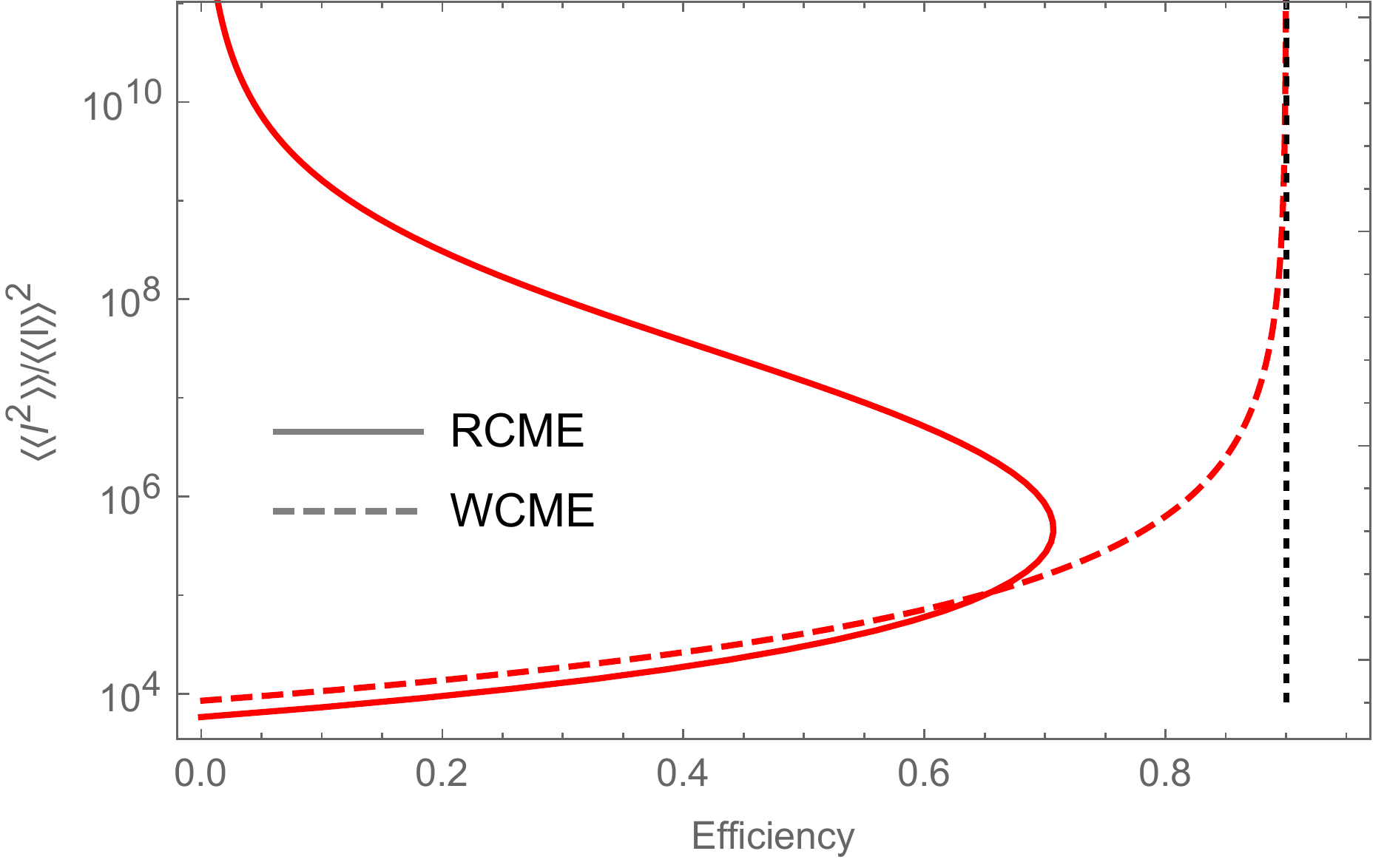}}}
     \caption{(a) Relative uncertainty as a function of reorganisation energy. %for a fixed temperature gradient and bias voltage. 
     (b,c) Parametric plots of relative uncertainty against power (b) and efficiency (c) generated by altering the bias voltage for a fixed temperature gradient and %$\lambda$ 
reorganisation energy. The black dashed line in (c) shows the Carnot efficiency. 
Other parameters used are $\Delta\beta_{\rm R}=2, \omega_0\beta_{\rm R}=\gamma\beta_{\rm R}=100$ and $\Gamma_{\rm L}\beta_{\rm R}=\Gamma_{\rm R}\beta_{\rm R}=0.1$.}
      \label{fig:PATThermoelectricLeadResourceUnc}
\end{figure}

\begin{figure*}[t]
     \subfloat[ ][$V \beta_{\rm el} =0.1$, $\beta_{\rm ph}=0.1\beta_{\rm el}$]{{\includegraphics[width=0.45\textwidth]{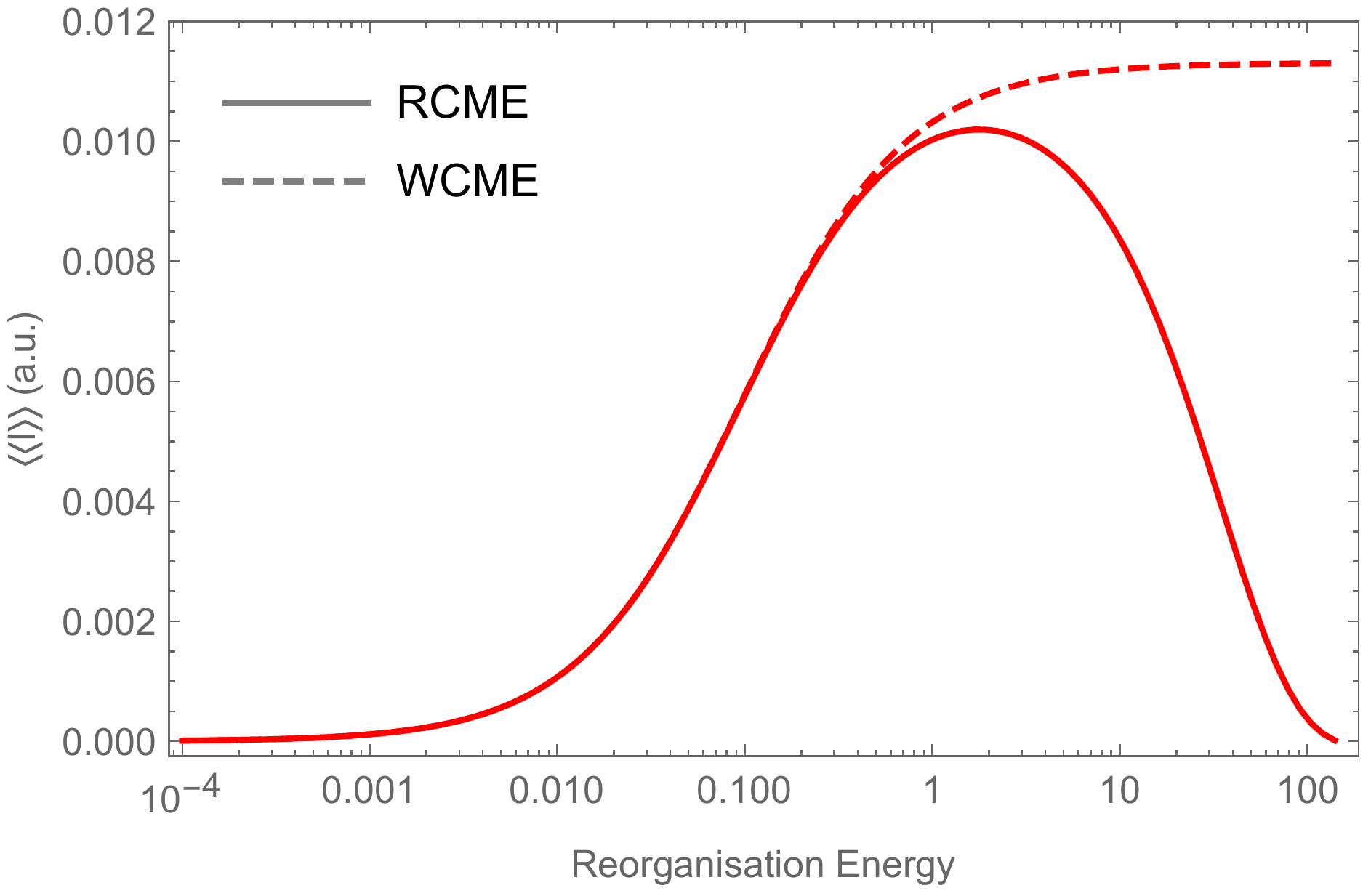}}}%
     \qquad    
    \subfloat[ ][$\lambda \beta_{\rm el}=3$, $\beta_{\rm ph}=0.1\beta_{\rm el}$]{{\includegraphics[width=0.45\textwidth]{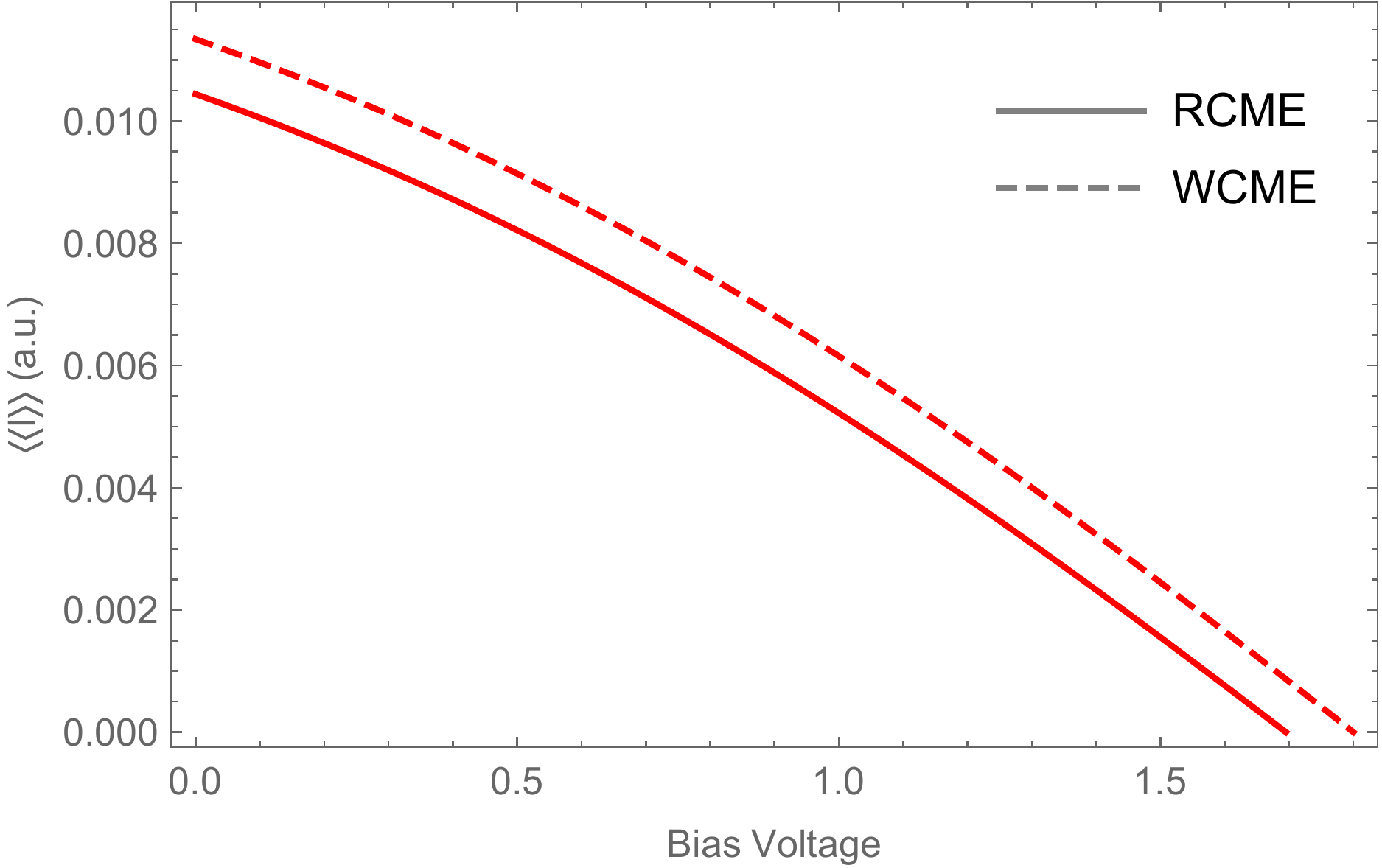}}}%
     \qquad \\ 
    \subfloat[ ][$\lambda \beta_{\rm el}=3$, $\beta_{\rm ph}=0.1\beta_{\rm el}$]{{\includegraphics[width=0.45\textwidth]{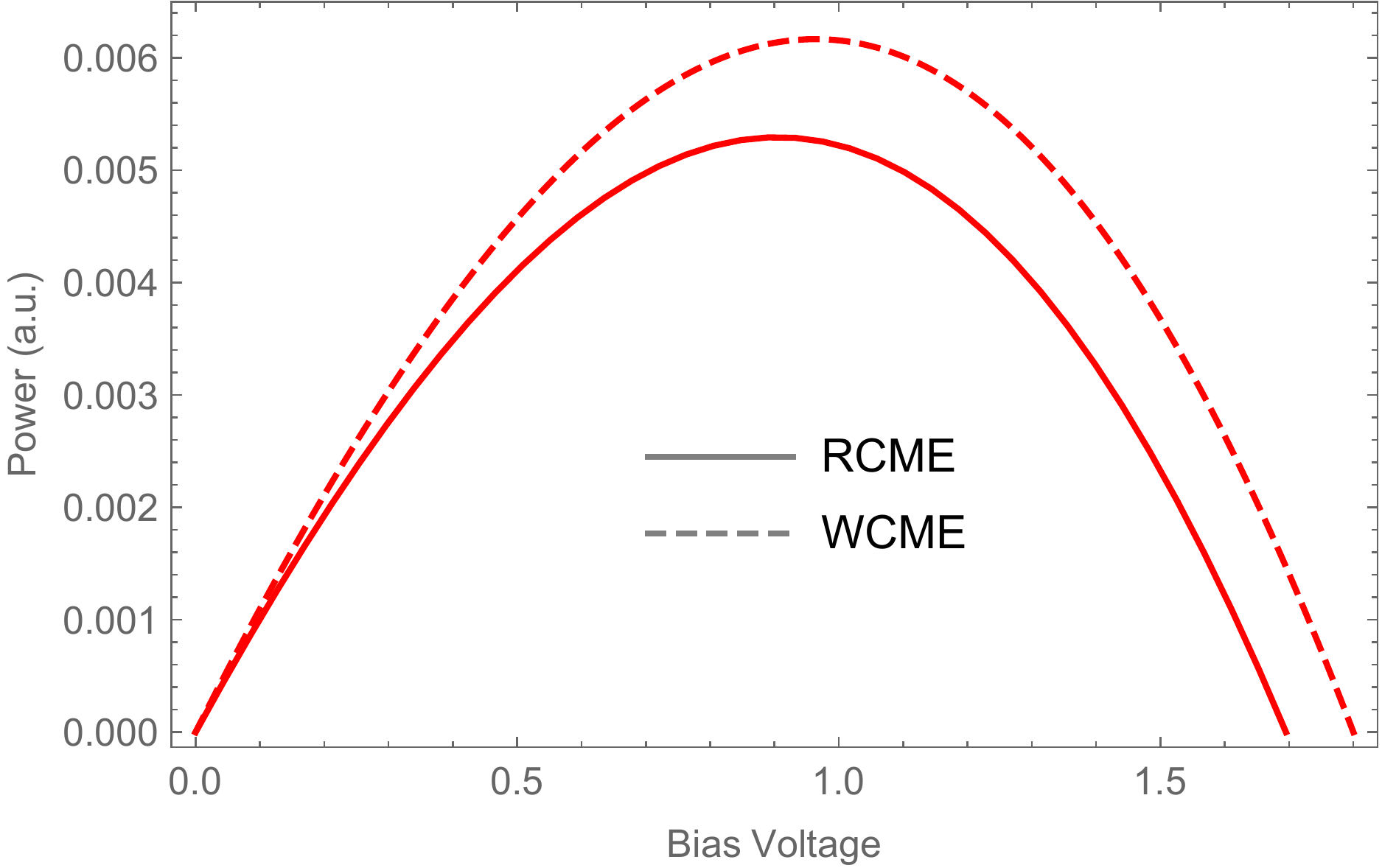}}}%
     \qquad 
     \subfloat[ ][%$\lambda \beta_{\rm el}=3$, $\beta_{\rm ph}=0.1\beta_{\rm el}$
     $V\beta_{\rm el} = 0.1$, $\beta_{\rm ph}=0.1\beta_{\rm el}$]{{\includegraphics[width=0.45\textwidth]{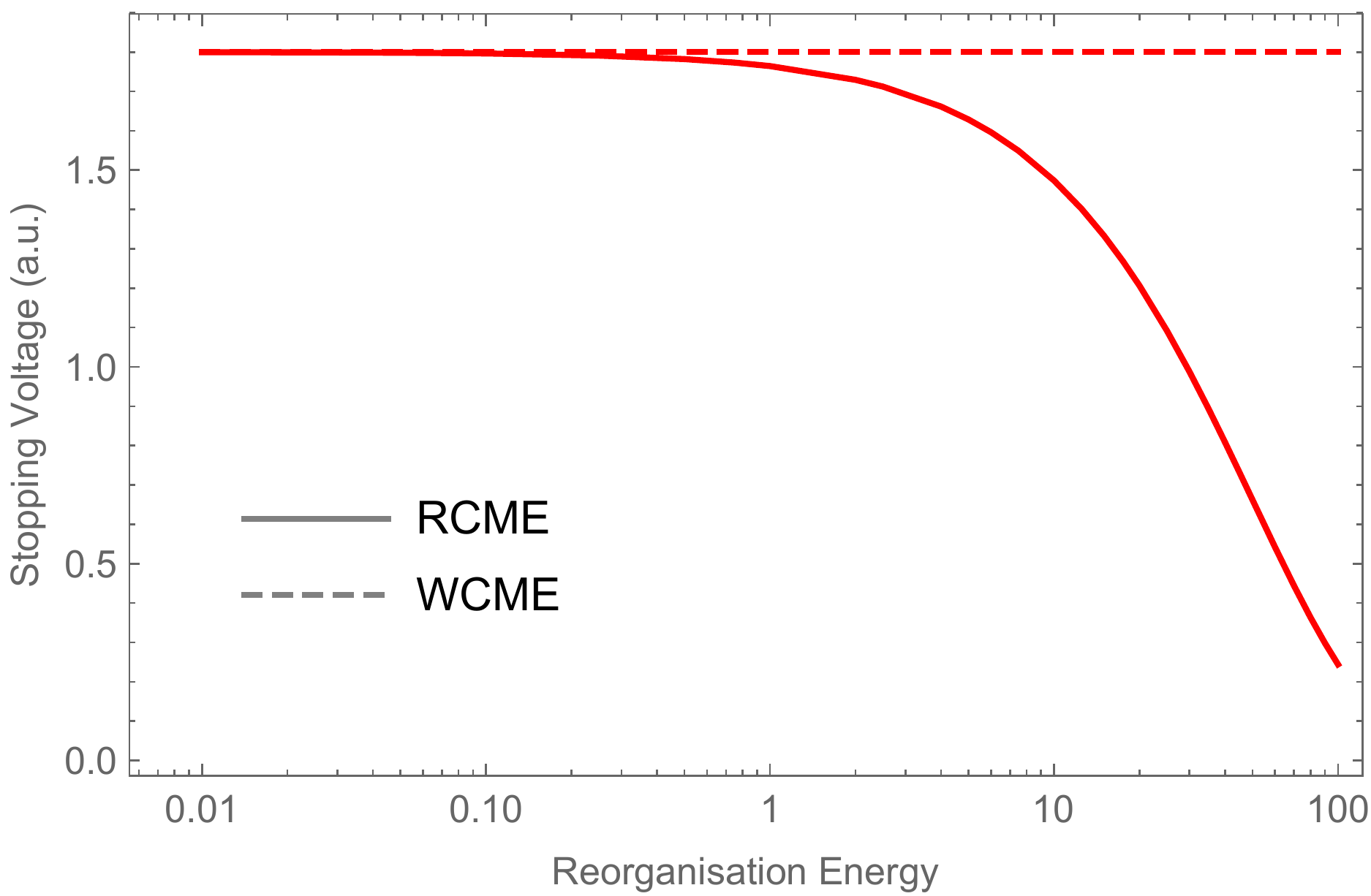}}}
     \caption{Current %(first order cumulant $\langle\langle I\rangle\rangle$) 
     as a function of (a) reorganisation energy $\lambda\beta_{\rm el}$ and (b) bias voltage $V\beta_{\rm el}$. (c) Power as a function of bias voltage. %for fixed temperature gradient and reorganisation energy. 
     (d) Stopping voltage as a function of reorganisation energy. %for fixed temperature gradient and bias voltage.
     Other parameters used are 
     $\Delta\beta_{\rm el}=2, \omega_0\beta_{\rm el}=\gamma\beta_{\rm el}=100$, and $\Gamma_{\rm L}\beta_{\rm el}=\Gamma_{\rm R}\beta_{\rm el}=0.1$.}
      \label{fig:PATThermoelectricPhononResourceCurrent}
\end{figure*}

These features of strong phonon coupling are also evident in the parametric plots of power and efficiency shown in Fig.~\ref{fig:PATThermoelectricLeadResourceEff}(d), which are calculated from the RCME by fixing the reorganisation energy and then sweeping through the entire bias window for which the system operates as a thermoelectric. 
As we increase the bias voltage the power and efficiency both increase until a maximum power is reached, at which point the power begins to decrease back to zero, while the efficiency increases to a maximum and then also returns to zero. 
Increasing the reorganisation energy results in an increase in the maximum power attainable, however the maximum efficiency is reduced as a result. This trade-off implies that a system with a given level of phonon coupling is suited to different roles (e.g.~maximising efficiency or power) depending on how weak or strong the reorganisation energy is. We also note that for increasing $\lambda$ the parametric plot maps  out a larger area due to the increased bias range over which the system acts as a thermoelectric. 

We now extend our analysis to consider higher cumulants, specifically the zero frequency noise $\langle\langle I^2\rangle \rangle$, and potential trade-offs between fluctuations and the thermoelectric power and efficiency. We are motivated %to do so due to
in part by recent advances in the study of thermodynamic uncertainty relations (TURs) 
\cite{PhysRevE.93.052145,PhysRevLett.116.120601,Seifert18,Liu19,Saryal19,Goold19,PhysRevLett.125.260604,PhysRevLett.126.210603}. Steady-state TURs provide cost-precision trade-off relations 
whereby the relative uncertainty in the current 
\begin{equation}
\upsilon=\frac{\langle\langle I^2\rangle \rangle}{\langle\langle I\rangle \rangle^2},
\end{equation}
is constrained by the entropy production, implying that for a more precise output with less noise, there will be a greater cost (higher entropy production). Conversely, if we want to minimise entropy production to maximise efficiency, this is expected to come at the cost of increasing the relative uncertainty. 

In Fig.~\ref{fig:PATThermoelectricLeadResourceUnc}(a) we plot the relative uncertainty as a function of phonon reorganisation energy for both the WCME and RCME. As the phonon coupling strength increases, the relative uncertainty decreases, corresponding to a reduction in the noise as compared to the increasing current (see Fig.~\ref{fig:PATThermoelectricLeadResourceCurrent}(a)). In particular, in the regime of enhanced current above $\lambda\beta_{\rm R}\sim1$ the RCME predicts a significant suppression of $\upsilon$. Here, power increases and fluctuations decrease, but the trade-off is that efficiency decreases as well. We see that the WCME overestimates $\upsilon$ in the same region, in some cases by over an order of magnitude, and it also fails to capture a very rapid increase in the relative uncertainty in the 
electron blockade regime for $\lambda\beta_{\rm R}>100$.

\begin{figure*}[t]
     \subfloat[ ][$\lambda \beta_{\rm el}=3$, $\beta_{\rm ph}=0.1\beta_{\rm el}$]{{\includegraphics[width=0.45\textwidth]{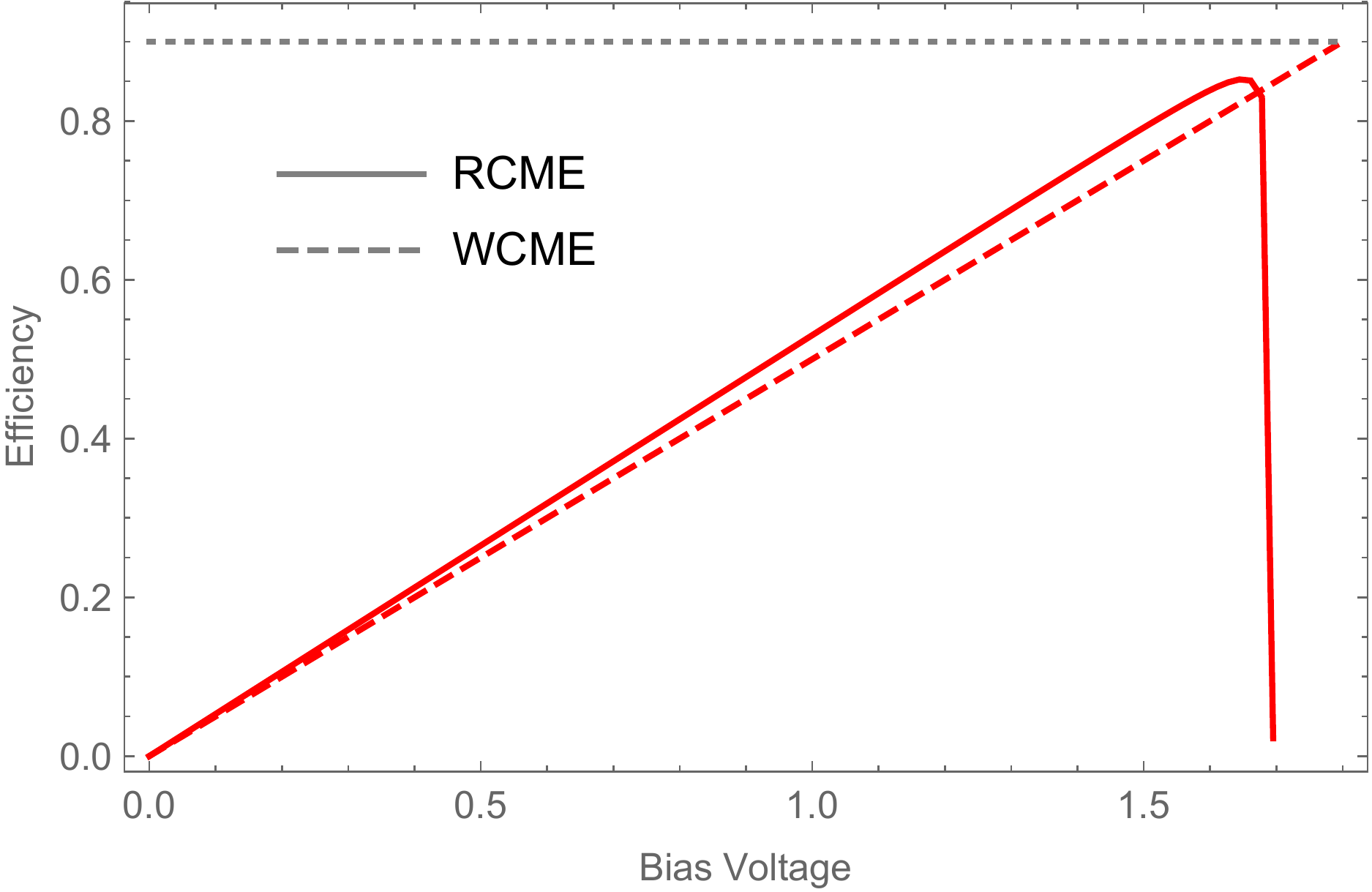}}}
     \qquad    
    \subfloat[ ][$V\beta_{\rm el} = 0.1$, $\beta_{\rm ph}=0.1\beta_{\rm el}$]{{\includegraphics[width=0.45\textwidth]{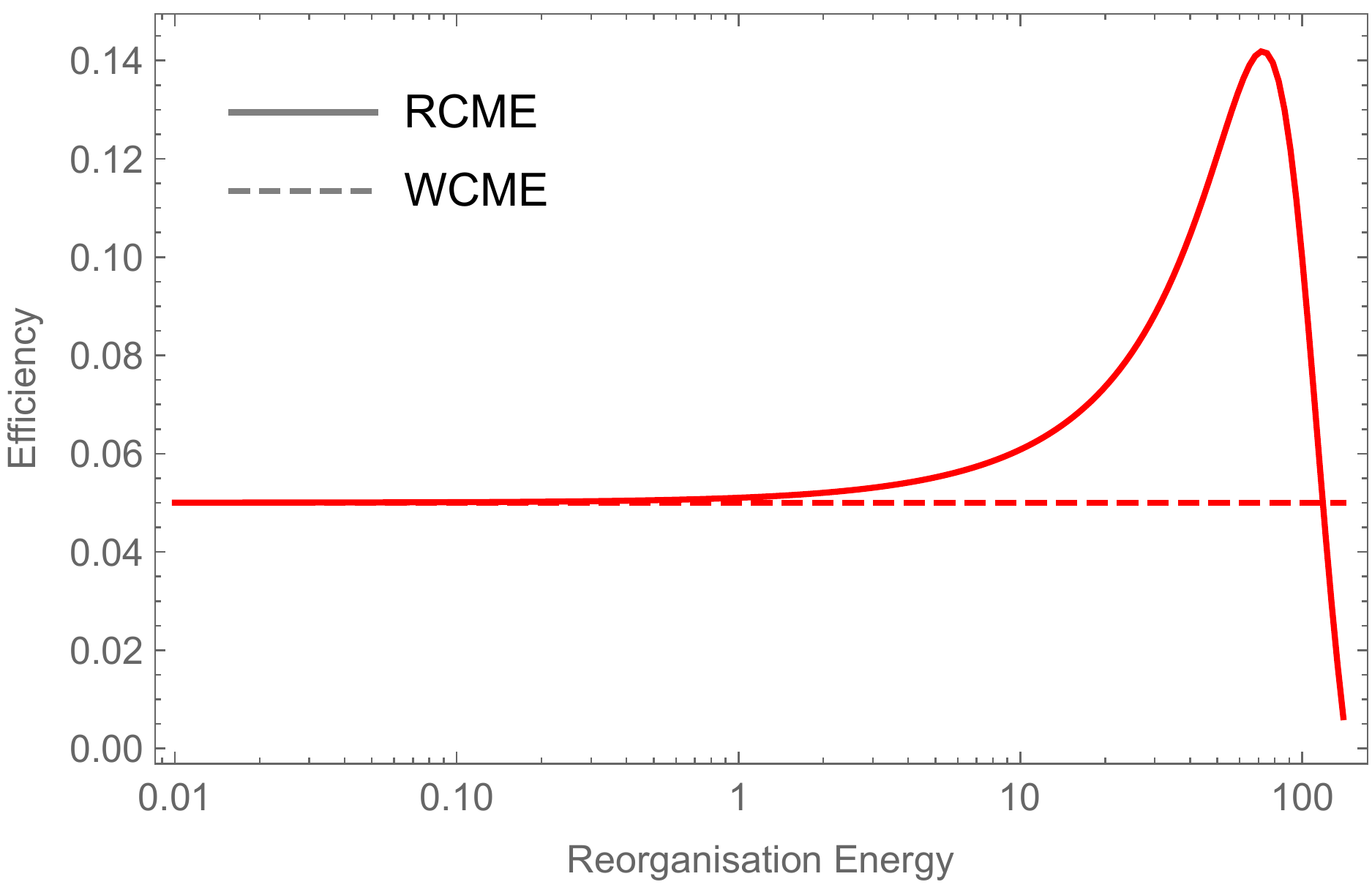}}}
     \qquad \\ 
    \subfloat[ ][$\beta_{\rm ph}=0.1\beta_{\rm el}$]{{\includegraphics[width=0.45\textwidth]{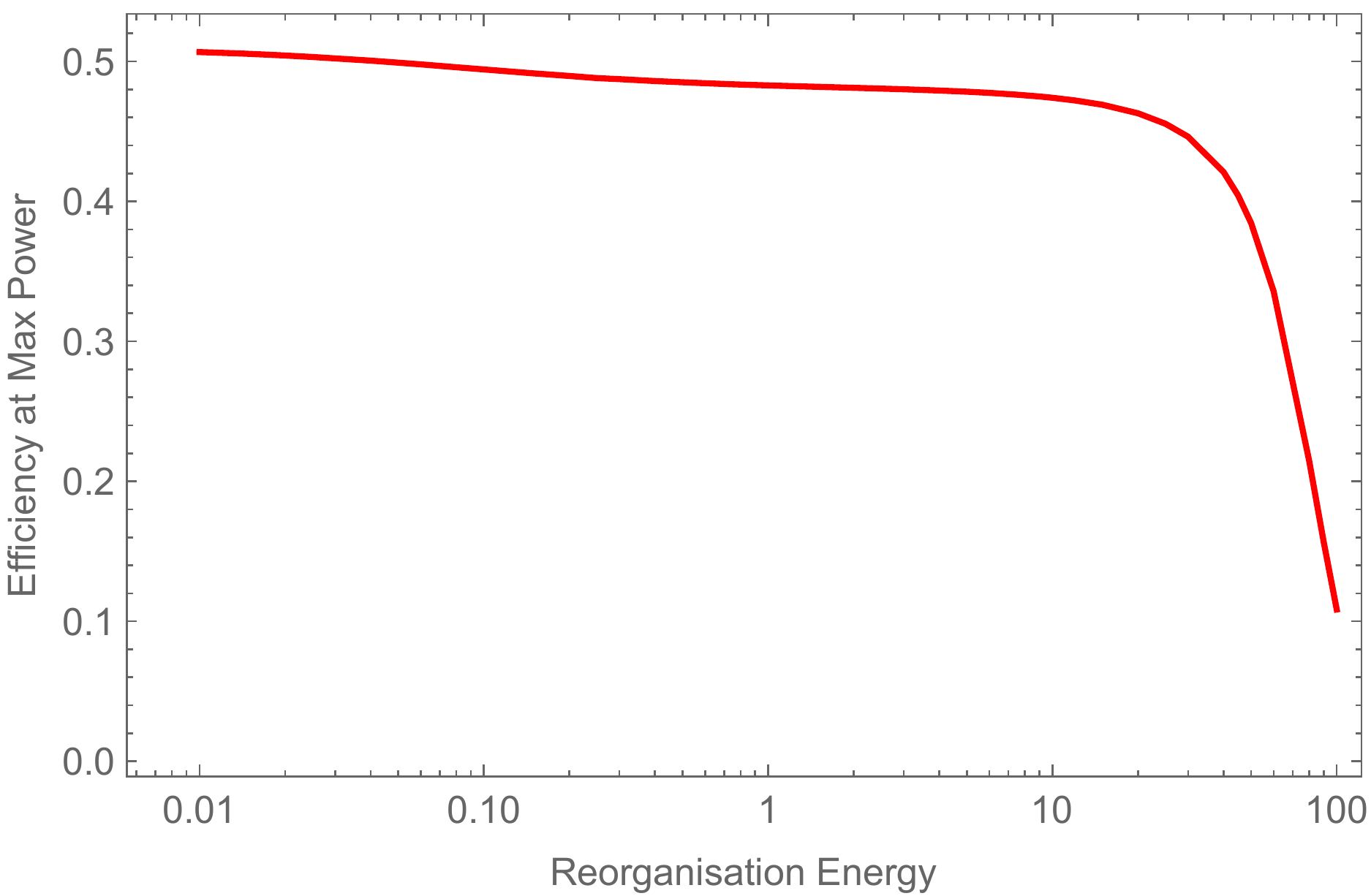}}}
     \qquad 
     \subfloat[ ][$\beta_{\rm ph}=0.1\beta_{\rm el}$]{{\includegraphics[width=0.45\textwidth]{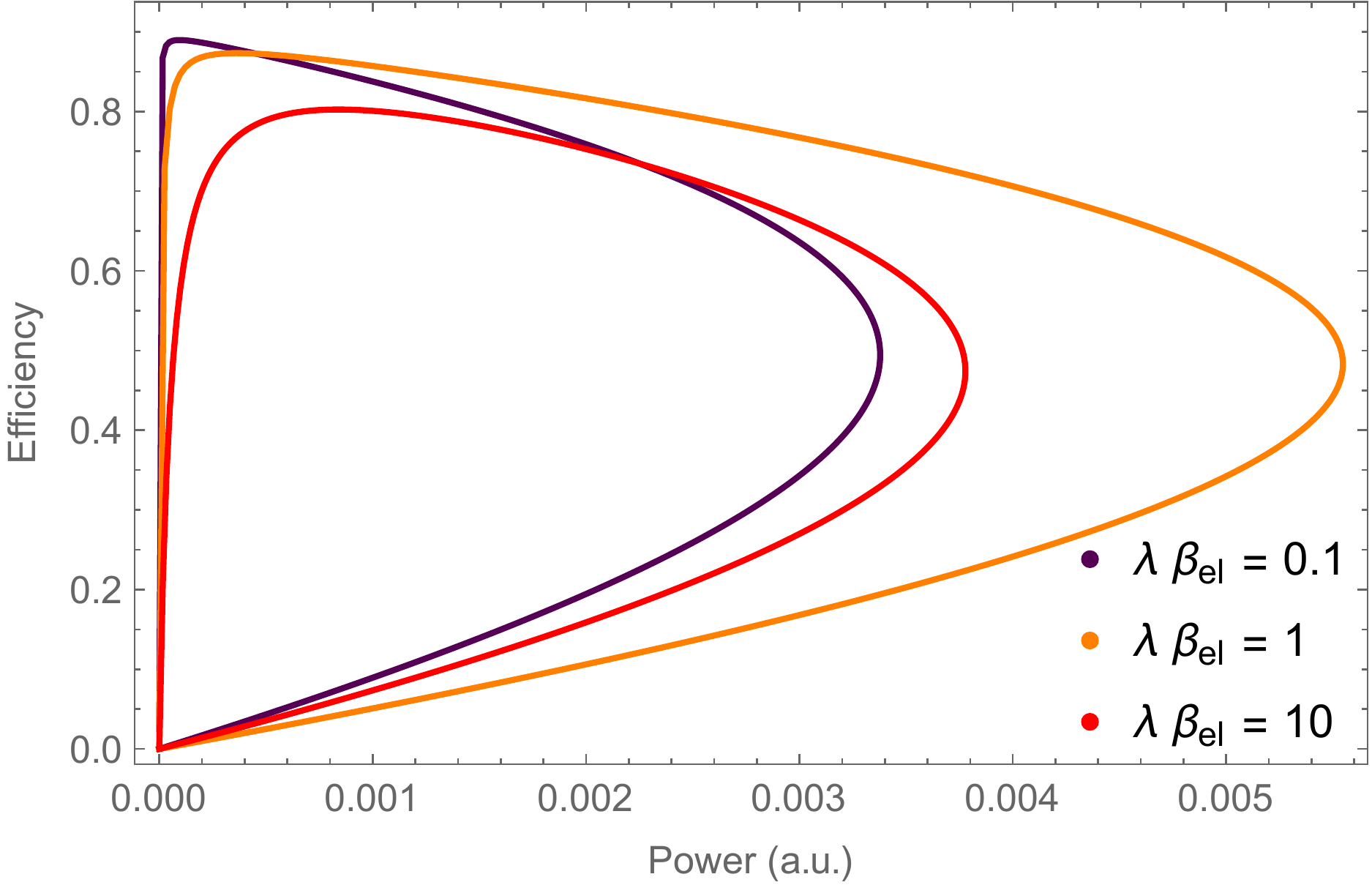}}}
     \caption{Efficiency against (a) bias voltage %, for fixed reorganisation energy and temperature gradient, 
     and (b) reorganisation energy. %, for fixed bias voltage and temperature gradient. 
     (c) Efficiency at maximum power against reorganisation energy. (d) Parametric plot of power and efficiency generated by altering the bias voltage for fixed temperature gradient and various reorganisation energies within the RCME treatment. Other parameters used are 
     $\Delta\beta_{\rm el}=2, \omega_0\beta_{\rm el}=\gamma\beta_{\rm el}=100$, and $\Gamma_{\rm L}\beta_{\rm el}=\Gamma_{\rm R}\beta_{\rm el}=0.1$.}
      \label{fig:PATPowerEfficiency}
\end{figure*}

\begin{figure}[t]
     \subfloat[ ][$V\beta_{\rm el}=0.1$, $\beta_{\rm ph}=0.1\beta_{\rm el}$]{{\includegraphics[width=0.45\textwidth]{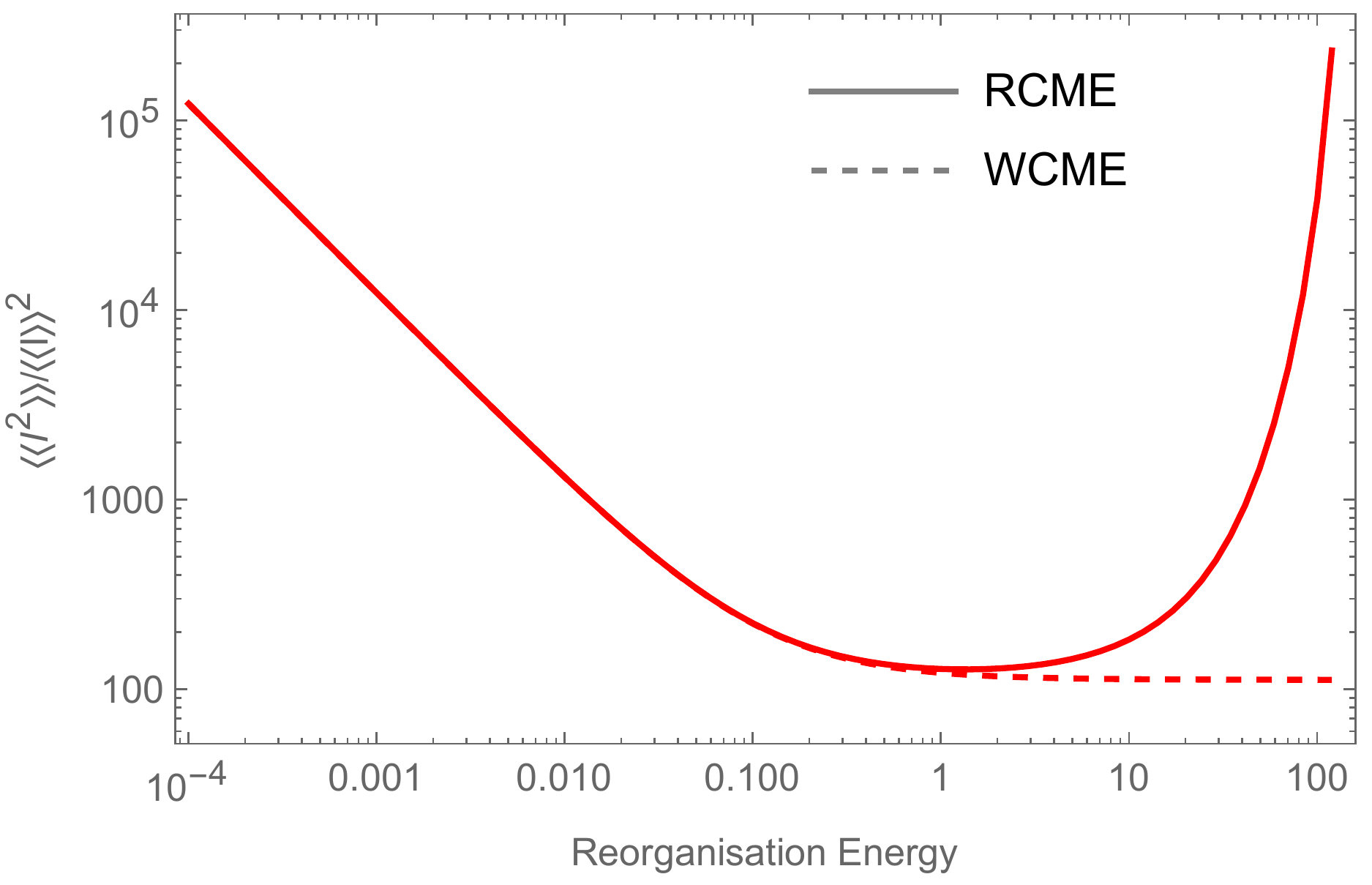}}}%
     \qquad    
    \subfloat[ ][$\lambda\beta_{\rm el}=1$, $\beta_{\rm ph}=0.1\beta_{\rm el}$]{{\includegraphics[width=0.45\textwidth]{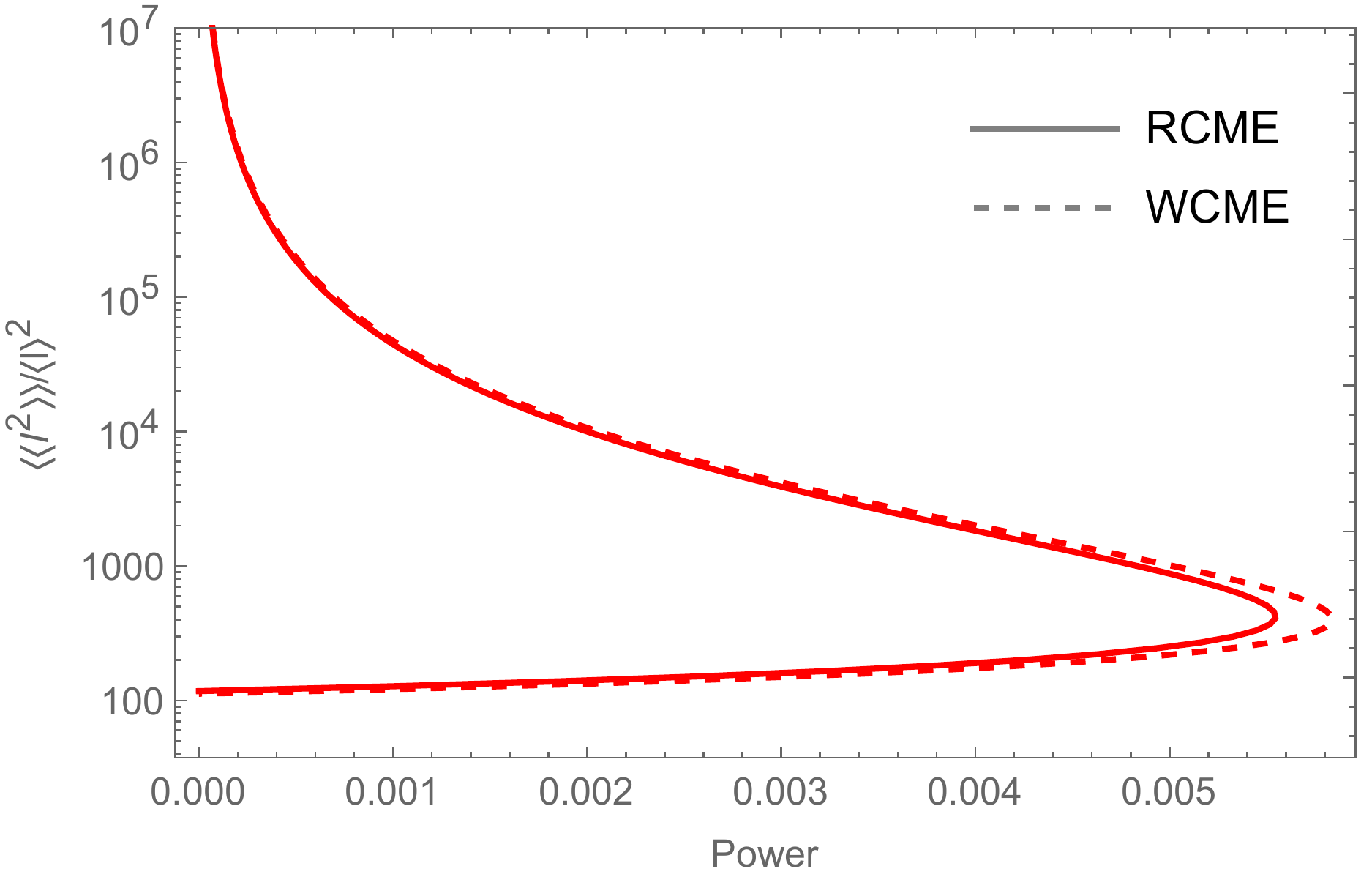}}}%
     \qquad \\ 
    \subfloat[ ][$\lambda\beta_{\rm el}=1$, $\beta_{\rm ph}=0.1\beta_{\rm el}$]{{\includegraphics[width=0.45\textwidth]{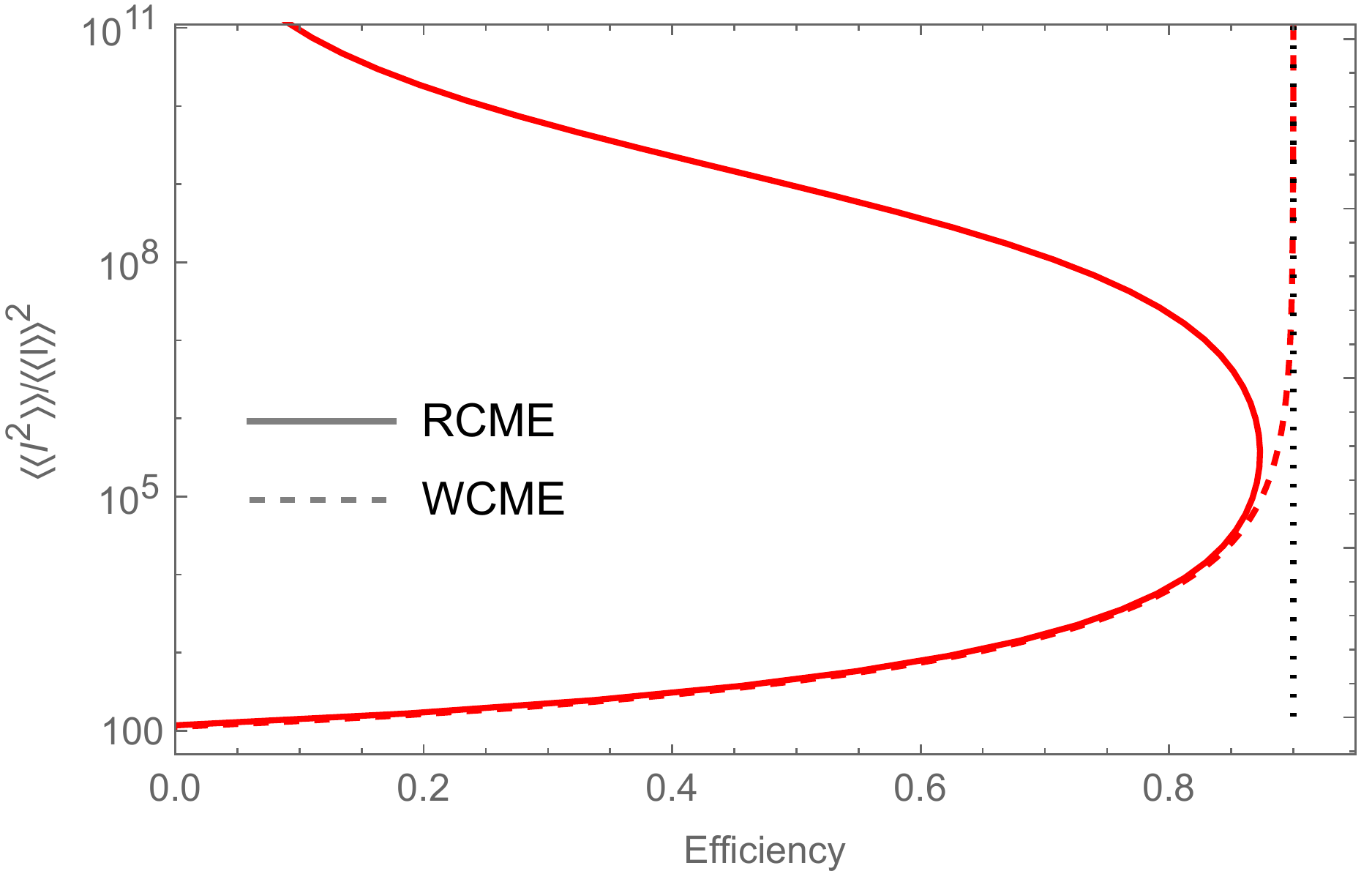}}}%
     \caption{(a) Relative uncertainty %($\langle\langle I^2\rangle\rangle/\langle\langle I\rangle\rangle^2$) 
     as a function of phonon reorganisation energy. (b,c) Parametric plots of relative uncertainty against power (b) and efficiency (c) calculated for changing bias voltage. The black dotted line in (c) shows the Carnot efficiency. Other parameters used are 
     $\Delta\beta_{\rm el}=2, \omega_0\beta_{\rm el}=\gamma\beta_{\rm el}=100$ and $\Gamma_{\rm L}\beta_{\rm el}=\Gamma_{\rm R}\beta_{\rm el}=0.1$.}
      \label{fig:PATThermoelectricPhononResourceNoise}
\end{figure}

We now fix the reorganisation energy to $\lambda\beta_{\rm R}=3$ and instead scan the voltage $V$ across the relevant bias window for which the system acts as a thermoelectric, producing parametric plots of the relative uncertainty against power (Fig.~\ref{fig:PATThermoelectricLeadResourceUnc}(b)) and efficiency (Fig.~\ref{fig:PATThermoelectricLeadResourceUnc}(c)). The qualitative features of the relative uncertainty against power plot (Fig.~\ref{fig:PATThermoelectricLeadResourceUnc}(b)) are similar for both the RCME and WCME, though absolute values can vary substantially. For smaller values of the bias voltage, power can be increased at a relatively modest increase in $\upsilon$, but once $V$ approaches the stopping voltage, the relative uncertainty begins to diverge rapidly and the current becomes extremely noisy. In this regime the weak-coupling analysis predicts the entropy production rate to fall to zero and the efficiency to approach the Carnot bound, see Fig.~\ref{fig:PATThermoelectricLeadResourceUnc}(c). The RCME prediction is qualitatively different, however. As the stopping voltage is approached in the RCME case $\upsilon$ still diverges, but the efficiency now falls to zero, consistent with our earlier analysis in Fig.~\ref{fig:PATThermoelectricLeadResourceEff}(a).

%It is in this regime that a weak-coupling analysis predicts the entropy production rate to fall to zero and the efficiency to approach the Carnot bound, see Fig.~\ref{fig:PATThermoelectricLeadResourceUnc}(c). The RCME prediction is qualitatively different, however, and not only does $\upsilon$ diverge, but the efficiency also falls to zero in this regime, consistent with our earlier analysis in Fig.~\ref{fig:PATThermoelectricLeadResourceEff}(a).

\section{Thermoelectric Regime II: Phonon Resource}\label{phononresource}

In the previous section we considered the left lead as the source of heat, with the phonons acting as a transport mediator as well as a heat sink. In this section we shall instead treat the phonons as a thermodynamic resource, allowing us to draw comparisons between the two regimes of operation. 
We thus set $\beta_{\rm ph}<\beta_{\rm el}$, where $\beta_{\rm el}=\beta_{\rm L}=\beta_{\rm R}$ is the inverse temperature of the leads. 
We retain the original chemical potential gradient, $\mu_{\rm R}>\mu_{\rm L}$, so it is only by exploiting the temperature gradient between the phonons and the leads that we can generate a positive current flow. 

\subsection{Thermoelectric performance}

Beginning with Fig.~\ref{fig:PATThermoelectricPhononResourceCurrent}(a) we see that the variation of the current with phonon coupling strength is qualitatively similar to the case when the left lead was the resource [Fig.~\ref{fig:PATThermoelectricLeadResourceCurrent}(a)], though now the WCME overestimates the current in comparison to the more accurate RCME.  
As a result of the dual role played by the phonon bath, as both thermodynamic resource and transport mediator, the RCME predicts a maximised current flow that is much larger and occurs for a much smaller reorganisation energy than in Fig.~\ref{fig:PATThermoelectricLeadResourceCurrent}(a), around $\lambda\beta_{\rm el}=1$. However, the multi-phonon processes captured by the RC mapping do not enhance the current here in comparison to the WCME, due to competition between phonon-mediated excitation and de-excitation within the manifold of augmented system states caused by the phonon bath now being at the highest temperature.

Similar trends are also reflected in the current and power behaviour with changes in bias voltage shown in Figs.~\ref{fig:PATThermoelectricPhononResourceCurrent}(b) and \ref{fig:PATThermoelectricPhononResourceCurrent}(c), respectively. 
Here we find that the stopping voltage is now reduced in the RCME treatment as compared to the WCME, and so the weak-coupling approximation overestimates the parameter range over which a thermoelectric current flows. 
In analogy with the left-lead resource regime, the stopping voltage is now given by $V_S = \Delta (\beta_{\rm el} - \beta_{\rm ph})/\beta_{\rm el}$ in the weak-coupling case. We noted previously that the RCME demonstrates an increased energy difference between the chemical potential of the left lead and the augmented system eigenvalues, thus increasing the stopping voltage when the left lead is the resource. In the phonon resource regime we find the reverse to be the case. It is now the energetic differences between the relevant tunnelling states that are important, and within the augmented system these are pushed closer together, resulting in 
a reduced stopping voltage as shown in Fig~\ref{fig:PATThermoelectricPhononResourceCurrent}(d). Once again, the WCME incorrectly predicts no variation in stopping voltage with changing reorganisation energy.

Figs.~\ref{fig:PATPowerEfficiency}(a) and (b) show the thermoelectric efficiency as a function of bias voltage and reorganisation energy, respectively. Again, the trend with varying bias voltage is similar to the left-lead resource regime, with the RCME predicting an increase in efficiency before a sudden drop to zero at the stopping voltage. The WCME generally underestimates the efficiency apart from in the regime beyond the point at which the RCME predicts the stopping voltage to occur. When fixing the bias voltage, as in Fig.~\ref{fig:PATPowerEfficiency}(b), an interesting trend is seen whereby the efficiency increases with increasing phonon reorganisation energy, before decreasing sharply for large coupling strengths. Despite this, the efficiency at maximum power, shown in Fig.~\ref{fig:PATPowerEfficiency}(c) decreases monotonically with increasing reorganisation energy. This is caused by a shift in the bias voltage that produces the maximum power output, which decreases for larger reorganisation energy, and so the efficiency does as well. In fact, considering the full bias range for which the system acts a a thermoelectric, as in Fig.~\ref{fig:PATPowerEfficiency}(d), reveals that the maximum obtainable efficiency is greater for smaller reorganisation energies. The maximum obtainable power does increase for larger $\lambda$, but the behaviour is non-monotonic, such that for $\lambda\beta_{\rm el}=10$ both the maximum power and efficiency that can be reached are lower than for $\lambda\beta_{\rm el}=1$.

Turning to current fluctuations in the phonon-resource regime, we plot the relative uncertainty ($\upsilon$) as a function of reorganisation energy in Fig.~\ref{fig:PATThermoelectricPhononResourceNoise}(a). The trend is similar to the case of a left-lead resource, with $\upsilon$ decreasing as the reorganisation energy increases and more current flows, until it reaches a minimum near the current maximum. For larger reorganisation energies the relative uncertainty increases rapidly primarily due to the suppression of current. As expected, the WCME only captures the correct trend up to a certain phonon coupling strength. Fixing the reorganisation energy and scanning the bias voltage we obtain parametric plots of relative uncertainty against power and efficiency, shown in Figs.~\ref{fig:PATThermoelectricPhononResourceNoise}(c) and (d), respectively. Again, these are qualitatively very similar to those of the left-lead resource regime shown in Figs.~\ref{fig:PATThermoelectricLeadResourceUnc}(b) and (c). For example, to reduce fluctuations smaller values of the bias voltage are favoured (in contrast to the conditions needed to maximise power or efficiency), and the RCME once more predicts that the Carnot efficiency cannot be reached, even for diverging relative uncertainty and vanishing power as the stopping voltage is approached. 

\subsection{Breakdown of additivity}

Finally, we shall %consider 
explore the breakdown of %the aRCME when used
environmental additivity within 
our thermoelectric model to highlight the importance of accurately capturing non-additive effects %non-additivity of 
for multiple environments beyond weak-coupling. This is the case even if only a single environment is strongly coupled to the system, 
as we demonstrate through use of the aRCME. The aRCME is an additive simplification of the RCME wherein  
the electron-phonon coupling is accurately modelled for strong reorganisation energies, but the lead couplings are treated phenomenologically, i.e.~they are derived assuming no alterations to the system caused by the strong electron-phonon coupling.

\begin{figure}[t]
     \subfloat[ ][$V\beta_{\rm el}=0.1$, $\beta_{\rm ph}=0.1\beta_{\rm el}$]{{\includegraphics[width=0.45\textwidth]{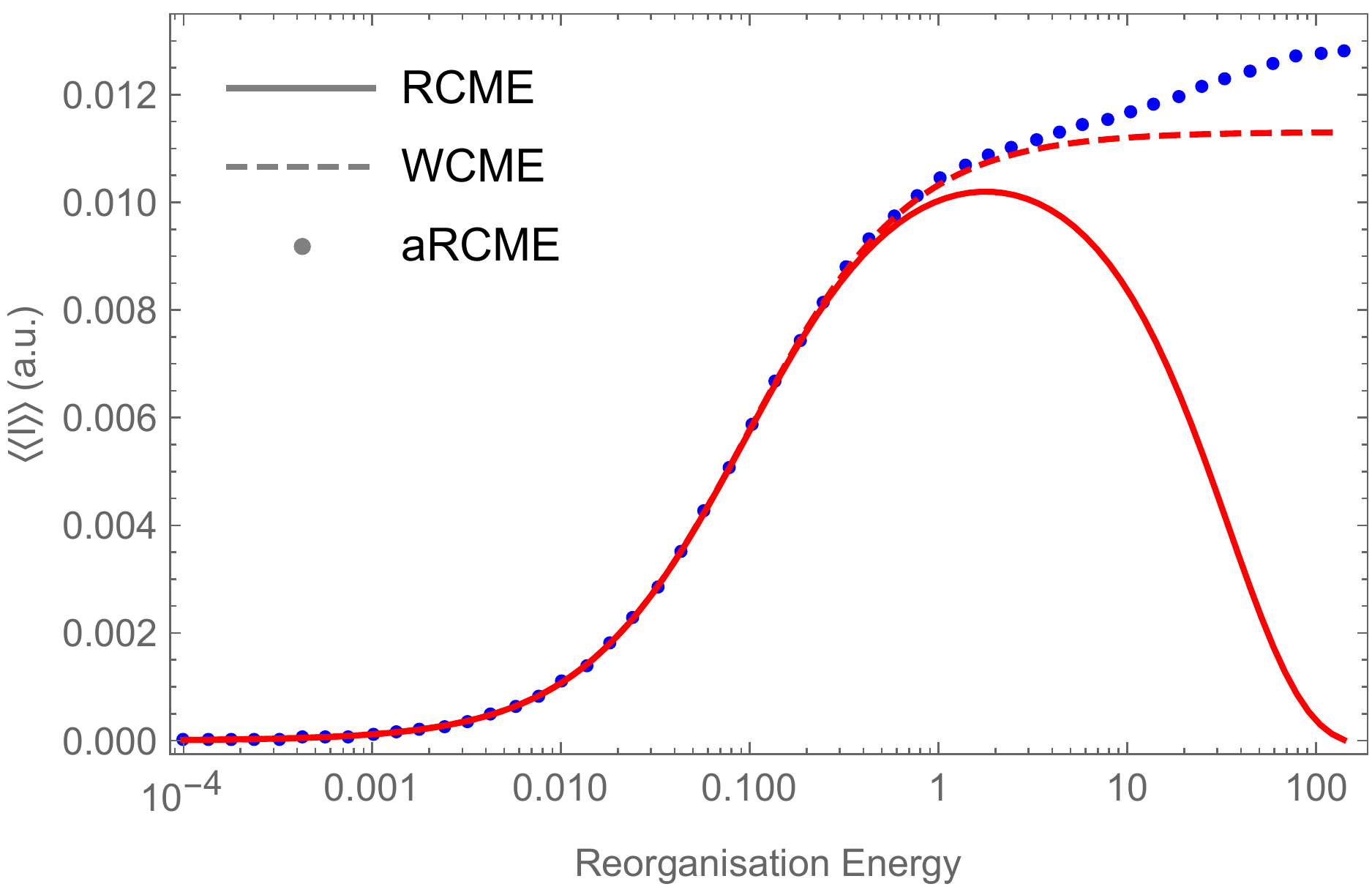}}}%
     \qquad
       \subfloat[ ][$V\beta_{\rm el}=0.1$, $\beta_{\rm ph}=0.1\beta_{\rm el}$]{{\includegraphics[width=0.45\textwidth]{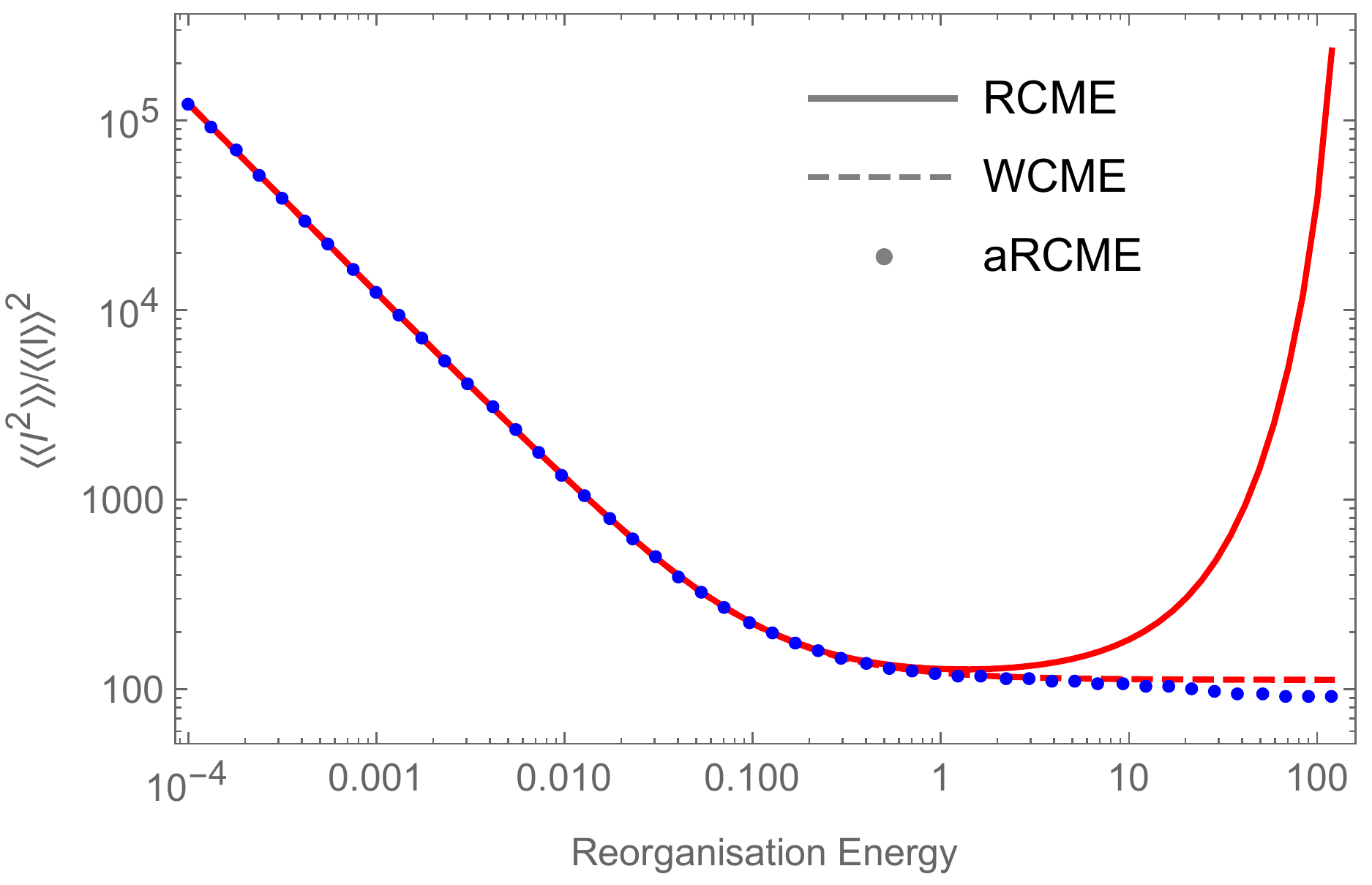}}}    
     \caption{Current flow (a) and relative uncertainty (b) against phonon reorganisation energy. Other parameters used are 
 $\Delta\beta_{\rm el}=2, \omega_0\beta_{\rm el}=\gamma\beta_{\rm el}=100$, and $\Gamma_{\rm L}\beta_{\rm el}=\Gamma_{\rm R}\beta_{\rm el}=0.1$.}
      \label{fig:PATThermoAdd}
\end{figure}

\begin{figure}[t]
     \subfloat[ ][$\lambda\beta_{\rm el}=5$, $\beta_{\rm ph}=0.1\beta_{\rm el}$]{{\includegraphics[width=0.45\textwidth]{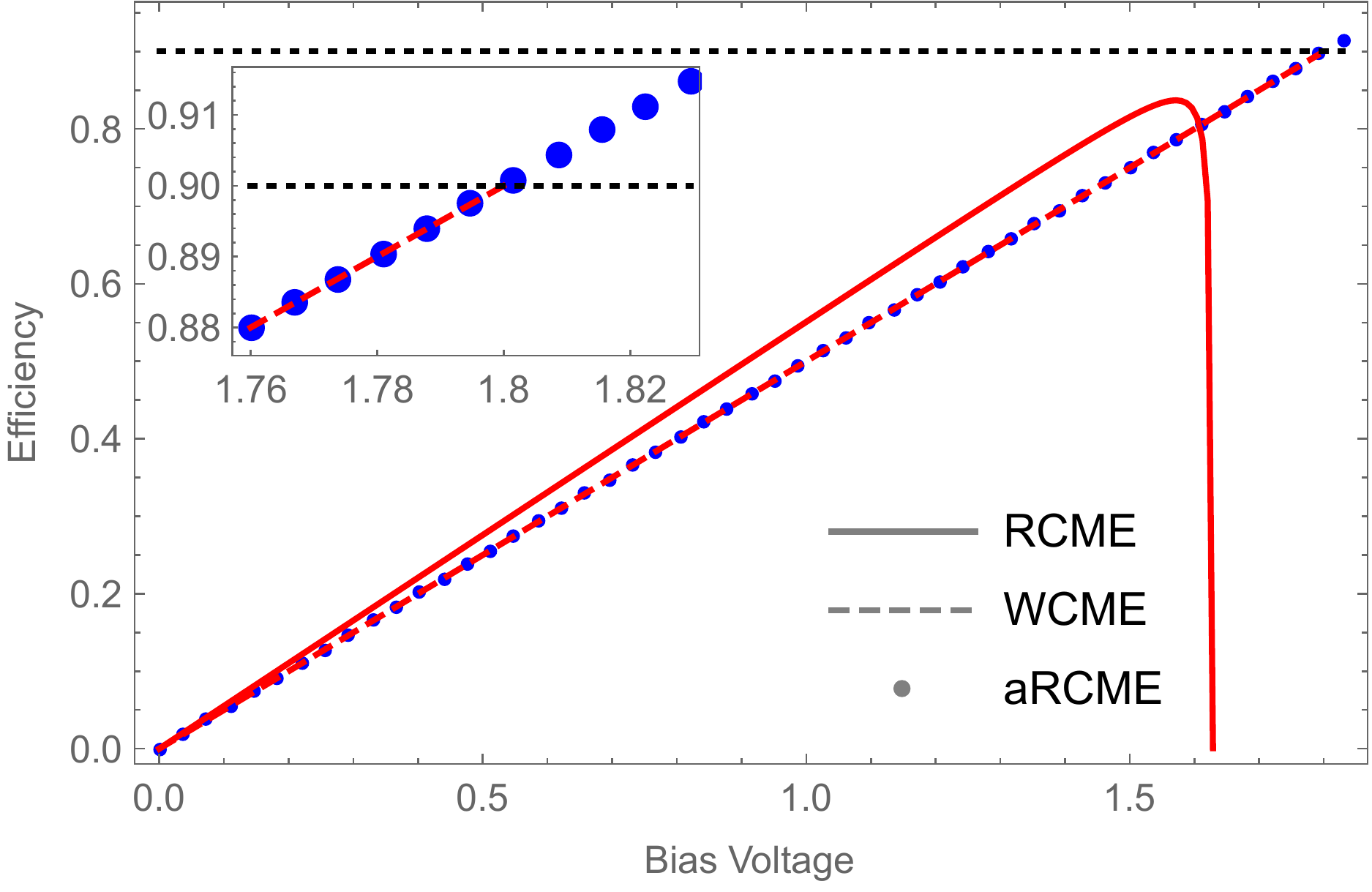}}}
     \caption{Efficiency against bias voltage. The black dashed lines indicate the Carnot efficiency and the inset shows a magnification for large bias voltage. Other parameters used are 
 $\Delta\beta_{\rm el}=2, \omega_0\beta_{\rm el}=\gamma\beta_{\rm el}=100$, and $\Gamma_{\rm L}\beta_{\rm el}=\Gamma_{\rm R}\beta_{\rm el}=0.1$.}
      \label{fig:PATThermoAddeff}
\end{figure}

In Fig.~\ref{fig:PATThermoAdd} we consider the current flow (a) and relative uncertainty (b) predicted by the aRCME for varying phonon reorganisation energy, 
and compare them with the previous results for the WCME and the full (non-additive) RCME in the phonon-resource regime. 
For the current flow 
we see that the aRCME completely misses the electron localisation predicted by the RCME at large reorganisation energies, and instead incorrectly predicts a continually %fact has an 
increasing current with larger phonon coupling strength. This is also reflected in the relative uncertainty, which simply decreases with increasing reorganisation energy. Thus, it is not only the quantitative, but also the qualitative behaviour predicted by the aRCME that is incorrect. This occurs due to the inconsistent application of the RC mapping within the aRCME, whereby the leads count on the original system transport channels rather than those of the mapped (augmented) system. In contrast to the augmented system states, there is no suppression of vibrational overlaps between the original system states and the empty state, meaning that the aRCME is then unable to capture the localisation regime. Furthermore, this implies that the power and heat flow definitions given by the aRCME are not thermodynamically consistent, as they are defined with respect to the original system states, yet the non-equilibrium steady-state is calculated with respect to the augmented system in the aRCME (albeit with phenomenological lead dissipators).

Though this failing is clear in the context of the RC treatment, more generally it highlights the importance of applying approximations consistently where multiple environments are concerned and at least one is to be treated beyond weak-coupling. A stark example of these effects is shown in Fig.~\ref{fig:PATThermoAddeff}, where the predictions for efficiency as a function of bias voltage match between the aRCME and WCME since they count on the same transport channels. Nevertheless, as the aRCME definitions also (incorrectly) give rise to an increased stopping voltage, it has the potential to breakdown completely, predicting unphysical thermoelectric efficiencies beyond the Carnot limit. 

\section{Conclusions}\label{conc}

We have combined the RCME and counting statistics techniques to treat strong phonon-coupling in thermoelectric nanojunction models, avoiding the limitations of the standard Born-Markov WCME. Interestingly, we found discrepancies between the two approaches not only at very strong phonon coupling, as would be expected, but also for relatively moderate phonon coupling strengths. For example, the WCME incorrectly predicts efficiencies and stopping voltages that are completely independent of the phonon reorganisation energy, highlighting the importance of an accurate representation of the electron-phonon coupling. We have also emphasised potential pitfalls of employing a phenomenological additive treatment of the influence of multiple baths when at least one couples strongly to the system. In the present context, this was shown to lead to an unphysical violation of the Carnot bound to the thermoelectric efficiency.

More generally, though there are common trends, the thermoelectric behaviour we have found appears also to be sensitive to the chosen parameter regimes, and whether strong coupling is beneficial or detrimental depends to a certain extent on which figure of merit is being optimised. Nevertheless, we do find that trade-offs seem to exist between power, efficiency, and fluctuations at strong coupling, and it would be interesting to explore whether this could be formalised in a more model-independent manner. 
In future work, we 
also plan to explore the interplay between strong lead and phonon couplings by extracting RCs not only for the vibrational degrees of freedom but also for the fermionic environments~\cite{PhysRevB.97.205405,RCReview}. 

{\em Acknowledgments.} This work was supported by the UK Engineering and Physical Sciences Research Council, grant no. EP/N008154/1.

%\bibliography{Bibliography.bib}

\end{document}